%% file: main.tex
\documentclass[11pt]{article}

\usepackage[final]{acl}

\usepackage{times}
\usepackage{latexsym}
\usepackage[T1]{fontenc}
\usepackage[utf8]{inputenc}
\usepackage{microtype}
\usepackage{inconsolata}

\usepackage{graphicx}
\usepackage{booktabs}
\usepackage{multirow}
\usepackage{amsmath}
\usepackage{amssymb}
\usepackage{xcolor}
\usepackage{xspace}
\usepackage{enumitem}
\usepackage{algorithm}
\usepackage{algpseudocode}
\usepackage{tikz}
\usetikzlibrary{shapes.geometric, arrows.meta, positioning, fit, backgrounds, calc}

\newcommand{\method}{SaP\xspace}
\newcommand{\methodlong}{Skill-as-Pseudocode\xspace}

\title{\methodlong: \\ Refactoring Skill Libraries to Pseudocode for LLM Agents}

\author{
  Xinze Li$^{1,3}$ \quad Yuhang Zang$^{3}$ \quad Yixin Cao$^{2\dagger}$ \quad Aixin Sun$^{1\dagger}$ \\[3pt]
  $^{1}$Nanyang Technological University, Singapore \quad
  $^{2}$Fudan University, Shanghai, China \\[1pt]
  $^{3}$Shanghai AI Laboratory, Shanghai, China \\[2pt]
  $^{\dagger}$Corresponding authors.
}

\hypersetup{
  pdftitle={Skill-as-Pseudocode: Refactoring Skill Libraries to Pseudocode for LLM Agents},
  pdfauthor={Xinze Li, Yuhang Zang, Yixin Cao, Aixin Sun},
  pdfsubject={Findings of EMNLP 2026},
  pdfkeywords={LLM agents, skill libraries, typed contracts, pseudocode, retrieval-augmented agents}
}

\begin{document}
\maketitle

\input{sections/abstract}
\input{sections/introduction}
\input{sections/related_work}
\input{sections/method}
\input{sections/experiments}
\input{sections/analysis}
\input{sections/conclusion}

\input{sections/limitations}
\input{sections/ethics}

\input{sections/acknowledgments}

\bibliography{references,custom}

\appendix
\input{sections/appendix}

\end{document}

%% file: sections/abstract.tex
\begin{abstract}
Markdown skill libraries for LLM agents ship as free-form prose,
forcing the agent to re-derive both the input schema and the
concrete invocation syntax on every retrieval. This produces a
``confused $\to$ re-retrieve $\to$ still confused'' loop: the agent
issues a partially-correct action, receives uninformative feedback,
and re-retrieves the same prose. We propose \methodlong{}
(\method{}), an automatic conversion of markdown skill libraries
into typed pseudocode with deterministic quality control. From each
cluster of similar procedural passages, \method{} extracts a typed
contract and filters it through a four-check deterministic verifier
(coverage, binding, replacement, risk). Promoted contracts are
inlined into a rewritten skill skeleton alongside restored action
templates, giving the agent two complementary signals: a typed
signature for \emph{what} a skill does and a concrete template for
\emph{how} to invoke it. On the ALFWorld unseen split
($134$ games, \texttt{gpt-4o-mini}, three seeds), \method{} wins $82 / 402$
paired games versus $47 / 402$ for the Graph-of-Skills (GoS)
baseline (pooled McNemar $p = 8.2 \times 10^{-5}$), at
$-22.8 \pm 6.4\%$ input tokens and $-14.5 \pm 4.1\%$ LLM calls per
game. A bundle-component ablation attributes the gain to the
\emph{pairing} of typed contracts with concrete action templates:
the contract alone falls below the prose baseline.
\end{abstract}

%% file: sections/introduction.tex
\section{Introduction}
\label{sec:intro}

Modern LLM-based agents \citep{yao2023react,schick2023toolformer}
rely on \emph{skill libraries}---collections of pre-authored
procedural documents the agent retrieves and follows at task time.
Typed catalogs such as OpenAPI specifications
\citep{openapi2024spec,song2023restgpt,qin2024toolllm,patil2024gorilla}
already expose machine-readable function signatures, but the more
common deployment surface for agent skills today is \emph{markdown}:
Anthropic's \texttt{SKILL.md} documents \citep{anthropic2025skills},
the Graph-of-Skills (GoS) library used as our experimental
substrate, and server-side MCP descriptions
\citep{anthropic2024mcp} all ship as free-form prose for
human and LLM readers.

\paragraph{Prose is the bottleneck.}
We argue, and our experiments support, that the prose
representation itself, not the absence of factoring, is the
dominant bottleneck when LLM agents consume markdown skill
libraries. A markdown skill document does not separate \emph{what}
the skill produces (its output schema) from \emph{how} the skill
is invoked (the concrete primitive actions the environment
accepts), and the two are often interleaved with prose hints,
caveats, and worked examples. At retrieval time the agent must
re-derive both on every call. Empirically this produces a recurring
\emph{retrieval-action feedback loop} on ALFWorld: the agent
issues a \texttt{SkillRequest}, reads back a long markdown body,
emits an action whose verb or argument is slightly off
(\texttt{put X in/on Y} rather than \texttt{move X to Y};
\texttt{heat mug} before holding the mug), the environment returns
``\texttt{Nothing happens.}'', and the agent re-retrieves looking
for the missing detail
(\autoref{sec:analysis:failures}). The same loop drives both task
failures and the per-game LLM-call cost.

\paragraph{Typed pseudocode is a better representation.}
Two pieces of information short-circuit the loop. (1) A typed
\emph{contract} that exposes the skill's trigger, input schema,
output schema, and pre/post-conditions tells the agent in one
glance \emph{what} it can do and \emph{what arguments it must
supply}; this is the same affordance that OpenAPI specs give
typed-routing agents. (2) The \emph{concrete action templates}
(``\texttt{go to \{recep\}}'', ``\texttt{heat \{obj\} with
\{appliance\}}'') that the contract abstracts over tell the agent
\emph{how} to actually invoke the underlying environment, in the
exact tokens the environment accepts. Together these two pieces
form a structured pseudocode spec: a typed signature paired with
template-level concrete syntax. Markdown skill libraries today
contain both pieces of information but only in scattered prose
form, and existing typed-routing methods
\citep{song2023restgpt,qin2024toolllm,patil2024gorilla} assume the
contract has already been hand-written upstream.

\paragraph{\methodlong{} (\method{}).}
We propose to \emph{automate the prose-to-pseudocode conversion}
of markdown skill libraries with deterministic quality
control.\footnote{Code and pre-built artifacts:
\url{https://github.com/InternLM/Skill-as-Pseudocode}.}
Each original markdown skill (which we call a \emph{parent skill}
because the rewrite layer inlines child contracts beneath it)
parses into a sequence of \emph{procedural units}---markdown
sections under a heading, each describing one sub-procedure~\citep{li2023takebreak}. Units
across parents with similar verbs, objects, and code linkage are
grouped into a \emph{candidate cluster} of passages hypothesised
to describe the same underlying procedure. For each cluster, \method{} (i)~drafts
a typed contract $\kappa$ with a single LLM call (Stage~3),
(ii)~runs a four-check deterministic verifier on $\kappa$ for
\textbf{Coverage}, \textbf{Binding}, \textbf{Replacement}, and
\textbf{Risk} (\autoref{sec:method:verifier}), and (iii)~rewrites
each parent's matching units into an \texttt{invoke($\kappa$, args)}
placeholder grounded by per-call argument bindings. Two LLM-aware
post-verifier passes---\textbf{Binding Extraction} (BE) and
\textbf{Rewrite Cleanup} (RC)---ground the placeholder per
call-site and remove residual prose conflicts; both are validated
by deterministic post-checks (contracts remain referenced; required
inputs have bindings).

At retrieval time the agent receives a \emph{substituted bundle}
combining the rewritten parent skeleton with
\texttt{invoke($\kappa$, args)} placeholders, the inlined typed
contract for each invoked $\kappa$, and the restored concrete
action templates the placeholder stands for---delivering the
\emph{what} (contract) and \emph{how} (template) signals in a
single response.

\paragraph{Cross-parent factoring is a secondary benefit, not the
central claim.} Contracts callable in multiple parents are promoted
to standalone child skills (factoring is real,
\autoref{sec:results:yield}), but the retrieval-pool ablation
(\autoref{sec:results:retrieval_ablation}) shows the agent's gain
comes from the parent-side substituted bundle, not the child
library: surfacing children as standalone retrieval results
drops reward by $27\%$.

\paragraph{Contributions.}
(1) A reframing of skill-library improvement for LLM agents as
\emph{prose-to-typed-pseudocode conversion with deterministic
quality control}, complementary to typed-routing work
(\autoref{sec:related}) that assumes contracts already exist.
(2) The \method{} pipeline: a four-check deterministic verifier
plus two LLM-aware post-verifier passes (Binding Extraction and
Rewrite Cleanup), each guarded by a deterministic post-check
(\autoref{sec:method}); the gates make the pipeline portable across
extraction-model families, with divergence absorbed as rejections
rather than corrupted contracts
(\autoref{sec:results:crossmodel}).
(3) A hierarchical-only retrieval interface
that delivers a substituted bundle of typed contract + concrete
action template per \texttt{invoke} placeholder
(\autoref{sec:method:retrieval}). (4) On ALFWorld $134$ games with
\texttt{gpt-4o-mini}, pooled across three seeds, \method{} obtains
higher reward \emph{and} lower token cost than the GoS retrieval
baseline simultaneously: $82 / 402$ vs.\ $47 / 402$ paired wins
($+74\%$ relative reward; pooled McNemar exact
$p=8.2 \times 10^{-5}$) at $-22.8\pm 6.4\%$ input, $-17.3\pm 5.1\%$
output, and $-14.5\pm 4.1\%$ LLM calls per game; a
bundle-component ablation (\autoref{sec:results:bundle_ablation})
attributes the gain to the contract+template \emph{pairing} rather
than to structure or brevity alone. All empirical claims concern
\emph{markdown} skill libraries; typed-API surfaces (OpenAPI, MCP)
are an architectural extension we describe but do not evaluate
(\autoref{app:cross-modality}).

%% file: sections/related_work.tex
\section{Related Work}
\label{sec:related}

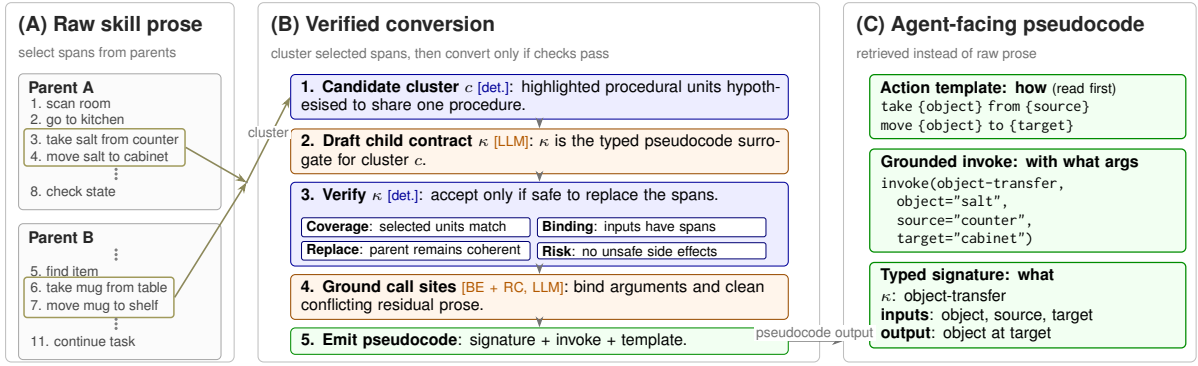
\begin{figure*}[!t]
  \centering
  \resizebox{0.99\textwidth}{!}{\input{figures/pipeline.tex}}
  \caption{\method{} as verified refactoring. Repeated prose spans in
  parent skills (A) feed a numbered, checked pipeline (B): form a
  candidate cluster $c$, draft a child contract $\kappa$, verify it,
  bind concrete arguments and clean residual conflicts, then emit an
  agent-facing pseudocode skill.
  The resulting representation (C) separates what the skill does
  (typed signature), what local arguments this parent supplies
  (grounded \texttt{invoke}), and how the environment is actually called
  (concrete action template).}
  \label{fig:pipeline}
\end{figure*}

\paragraph{Typed routing assumes the contract exists.}
RestGPT \citep{song2023restgpt}, ToolLLM \citep{qin2024toolllm},
Gorilla \citep{patil2024gorilla}, API-Bank \citep{li2023apibank}, and
the broader tool-augmented-agent paradigm
\citep{karpas2022mrkl,yao2023react,schick2023toolformer} together
with integration surfaces (Anthropic Skills
\citep{anthropic2025skills}, MCP \citep{anthropic2024mcp}, OpenAPI
\citep{openapi2024spec}) take a typed catalog as input and solve the
agent-side problem of \emph{selecting} and \emph{routing} calls;
\method{} is the complementary \emph{producer} side, recovering the
typed contract from prose markdown.

\paragraph{Skill-library maintenance and typed skill graphs.}
Concurrent work on library-level skill management includes
\textbf{SkillOps} \citep{skillops2026}, which represents each
skill as a typed contract and diagnoses library health
(redundancy, compatibility, missing validator) to drive
\texttt{merge}/\texttt{repair}/\texttt{retire}/\texttt{add\_adapter}
operations, and \textbf{GraSP} \citep{grasp2025}, which compiles
flat skill sets into a typed DAG for task-time composition. Both
operate on typed-contract representations \emph{after} the
contracts exist. \method{} addresses the upstream representation
gap: converting free-form prose skills into typed pseudocode
contracts plus inlined concrete action templates---the producer
side of the same supply chain. \textbf{SkillRet}
\citep{skillret2025} is a large-scale skill-retrieval benchmark
(skill selection, not skill content).

\paragraph{Library compression vs.\ verified refactoring.}
DreamCoder \citep{ellis2021dreamcoder}, Stitch
\citep{bowers2023stitch}, LAPS \citep{wong2021laps}, LILO
\citep{grand2024lilo}, and Agent Workflow Memory
\citep{wang2024agentworkflowmemory} extract reusable units from
$\lambda$-calculus or prose by MDL / reuse frequency. \method{}
differs in two ways: target is a typed pseudocode contract, and
recurrence only proposes candidates---callability is a separate
deterministic check, alongside self-correction
\citep{madaan2023selfrefine,gou2024critic} and process verification
\citep{lightman2024verify,uesato2022process} but with a deterministic
verifier. Trajectory-grown libraries (Voyager
\citep{wang2023voyager}, Reflexion \citep{shinn2023reflexion}, BOSS
\citep{zhang2024boss}, AutoManual \citep{chen2024automanual}, LATS
\citep{zhou2024lats}, SkillWeaver \citep{skillweaver2025}) require
successful executions; we assume a static library. Software refactoring
\citep{fowler1999refactoring,roy2007clone,shirafuji2023refactoring}
inspired the framing, the \emph{risk} check addresses tool-call
safety
\citep{ruan2024toolemu,liu2024promptinject,debenedetti2024agentdojo},
and Graph-of-Skills \citep{liu2026gos} (our retrieval baseline) is
within RAG \citep{lewis2020rag,zhou2023docprompting}; benchmarks are
ALFWorld
\citep{shridhar2021alfworld,ahn2022saycan,singh2023progprompt} and
SkillsBench \citep{benchflow2026skillsbench}.

%% file: figures/pipeline.tex
%

\begin{tikzpicture}[
    font=\sffamily,
    >={Stealth[length=1.7mm,width=1.2mm]},
    every node/.style={inner sep=0pt},
    panel/.style={
      draw=black!25, line width=0.45pt, rounded corners=2.5pt,
      fill=white,
    },
    title/.style={
      font=\sffamily\small\bfseries, text=black!85,
      anchor=north west,
    },
    subtitle/.style={
      font=\sffamily\tiny, text=black!55,
      anchor=north west,
    },
    parentbox/.style={
      draw=black!35, fill=gray!3, rounded corners=2pt, line width=0.4pt,
      minimum width=31mm, minimum height=20.8mm,
    },
    parenttitle/.style={
      font=\sffamily\scriptsize\bfseries, text=black!85,
      anchor=north west,
    },
    parentline/.style={
      font=\sffamily\tiny, text=black!72,
      anchor=west,
    },
    parentellipsis/.style={
      font=\sffamily\scriptsize, text=black!48,
      anchor=west,
    },
    selectionbox/.style={
      draw=yellow!55!black, rounded corners=1.4pt, line width=0.65pt,
      inner xsep=0.9mm, inner ysep=0.65mm,
    },
    stepbox/.style={
      draw=black!35, fill=gray!3, rounded corners=2pt, line width=0.45pt,
      text width=73mm, align=left,
      font=\sffamily\scriptsize,
      inner xsep=1.6mm, inner ysep=0.85mm,
    },
    llmstep/.style={
      stepbox, draw=orange!60!black, fill=orange!8,
    },
    detstep/.style={
      stepbox, draw=blue!60!black, fill=blue!7,
    },
    emitstep/.style={
      stepbox, draw=green!55!black, fill=green!5,
    },
    checkcell/.style={
      draw=blue!45!black, fill=white, line width=0.35pt,
      rounded corners=1pt,
      text width=33.5mm, align=left,
      font=\sffamily\tiny,
      inner xsep=0.75mm, inner ysep=0.45mm,
    },
    outcard/.style={
      draw=green!55!black, fill=green!5, rounded corners=2pt,
      line width=0.55pt,
      text width=45mm, align=left,
      font=\sffamily\scriptsize,
      inner xsep=1.7mm, inner ysep=0.95mm,
    },
    arrow/.style={->, draw=black!55, line width=0.5pt},
    spanarrow/.style={->, draw=yellow!45!black, line width=0.7pt},
    label/.style={
      font=\sffamily\tiny, text=black!55,
      fill=white, inner xsep=1pt, inner ysep=0.5pt,
    },
  ]

  \def\panelH{55mm}
  \def\panAW{35mm}
  \def\panBW{86mm}
  \def\panCW{54mm}
  \def\gap{3.5mm}

  \node[panel, anchor=north west, minimum width=\panAW, minimum height=\panelH]
    (inputPanel) at (0,0) {};
  \node[panel, anchor=north west, minimum width=\panBW, minimum height=\panelH]
    (convertPanel) at ($(inputPanel.north east)+(\gap,0)$) {};
  \node[panel, anchor=north west, minimum width=\panCW, minimum height=\panelH]
    (outputPanel) at ($(convertPanel.north east)+(\gap,0)$) {};

  \node[title] at ($(inputPanel.north west)+(2mm,-2mm)$)
    {(A) Raw skill prose};
  \node[subtitle] at ($(inputPanel.north west)+(2mm,-6.8mm)$)
    {select spans from parents};

  \node[parentbox, anchor=north west] (cardA)
    at ($(inputPanel.north west)+(2mm,-10.9mm)$) {};
  \node[parenttitle] at ($(cardA.north west)+(1.5mm,-1.2mm)$) {Parent A};
  \node[parentline] at ($(cardA.north west)+(1.8mm,-4.6mm)$) {1. scan room};
  \node[parentline] at ($(cardA.north west)+(1.8mm,-7.3mm)$) {2. go to kitchen};
  \node[parentline] (selA1) at ($(cardA.north west)+(1.8mm,-10.0mm)$)
    {3. take salt from counter};
  \node[parentline] (selA2) at ($(cardA.north west)+(1.8mm,-12.7mm)$)
    {4. move salt to cabinet};
  \node[selectionbox, fit=(selA1)(selA2)] (selA) {};
  \fill[black!50] ($(cardA.north west)+(15.0mm,-14.7mm)$) circle[radius=0.24mm];
  \fill[black!50] ($(cardA.north west)+(15.0mm,-15.4mm)$) circle[radius=0.24mm];
  \fill[black!50] ($(cardA.north west)+(15.0mm,-16.1mm)$) circle[radius=0.24mm];
  \node[parentline] at ($(cardA.north west)+(1.8mm,-18.0mm)$) {8. check state};

  \node[parentbox, anchor=north west] (cardB)
    at ($(cardA.south west)+(0,-2mm)$) {};
  \node[parenttitle] at ($(cardB.north west)+(1.5mm,-1.2mm)$) {Parent B};
  \fill[black!50] ($(cardB.north west)+(15.0mm,-4.1mm)$) circle[radius=0.24mm];
  \fill[black!50] ($(cardB.north west)+(15.0mm,-4.8mm)$) circle[radius=0.24mm];
  \fill[black!50] ($(cardB.north west)+(15.0mm,-5.5mm)$) circle[radius=0.24mm];
  \node[parentline] at ($(cardB.north west)+(1.8mm,-7.3mm)$) {5. find item};
  \node[parentline] (selB1) at ($(cardB.north west)+(1.8mm,-10.0mm)$)
    {6. take mug from table};
  \node[parentline] (selB2) at ($(cardB.north west)+(1.8mm,-12.7mm)$)
    {7. move mug to shelf};
  \node[selectionbox, fit=(selB1)(selB2)] (selB) {};
  \fill[black!50] ($(cardB.north west)+(15.0mm,-14.7mm)$) circle[radius=0.24mm];
  \fill[black!50] ($(cardB.north west)+(15.0mm,-15.4mm)$) circle[radius=0.24mm];
  \fill[black!50] ($(cardB.north west)+(15.0mm,-16.1mm)$) circle[radius=0.24mm];
  \node[parentline] at ($(cardB.north west)+(1.8mm,-18.0mm)$) {11. continue task};

  \node[title] at ($(convertPanel.north west)+(2mm,-2mm)$)
    {(B) Verified conversion};
  \node[subtitle] at ($(convertPanel.north west)+(2mm,-7mm)$)
    {cluster selected spans, then convert only if checks pass};

  \node[detstep, anchor=north west] (step1)
    at ($(convertPanel.north west)+(5mm,-11mm)$) {%
      \textbf{1. Candidate cluster $c$} {\tiny\textcolor{blue!60!black}{[det.]}}:
      highlighted procedural units hypothesised to share one procedure.
    };

  \node[llmstep, below=1.35mm of step1] (step2) {%
      \textbf{2. Draft child contract $\kappa$} {\tiny\textcolor{orange!70!black}{[LLM]}}:
      $\kappa$ is the typed pseudocode surrogate for cluster $c$.
    };

  \node[detstep, below=1.15mm of step2] (step3) {%
      \textbf{3. Verify $\kappa$} {\tiny\textcolor{blue!60!black}{[det.]}}:
      accept only if safe to replace the spans.\\[-0.1mm]
      \begin{tikzpicture}[baseline]
        \node[checkcell] (cov) {\textbf{Coverage}: selected units match};
        \node[checkcell, right=1mm of cov] (bind)
          {\textbf{Binding}: inputs have spans};
        \node[checkcell, below=0.6mm of cov] (rep)
          {\textbf{Replace}: parent remains coherent};
        \node[checkcell, right=1mm of rep] (risk)
          {\textbf{Risk}: no unsafe side effects};
      \end{tikzpicture}
    };

  \node[llmstep, below=1.15mm of step3] (step4) {%
      \textbf{4. Ground call sites} {\tiny\textcolor{orange!70!black}{[BE + RC, LLM]}}:
      bind arguments and clean conflicting residual prose.
    };

  \node[emitstep, below=1.15mm of step4] (step5) {%
      \textbf{5. Emit pseudocode}: signature + invoke + template.
    };

  \draw[arrow] (step1.south) -- (step2.north);
  \draw[arrow] (step2.south) -- (step3.north);
  \draw[arrow] (step3.south) -- (step4.north);
  \draw[arrow] (step4.south) -- (step5.north);

  \coordinate (merge) at ($(inputPanel.east)!0.52!(convertPanel.west)$ |- step1.west);
  \fill[black!45] (merge) circle[radius=0.55pt];
  \draw[spanarrow] (selA.east) -- (merge);
  \draw[spanarrow] (selB.east) -- (merge);
  \draw[spanarrow] (merge) -- (step1.west)
    node[label, midway, above]{cluster};

  \node[title] at ($(outputPanel.north west)+(2mm,-2mm)$)
    {(C) Agent-facing pseudocode};
  \node[subtitle] at ($(outputPanel.north west)+(2mm,-7mm)$)
    {retrieved instead of raw prose};

  \node[outcard, anchor=north west] (template)
    at ($(outputPanel.north west)+(4mm,-11mm)$) {%
      \textbf{Action template: how} \tiny(read first)\\[0.4mm]
      {\ttfamily\scriptsize take \{object\} from \{source\}}\\
      {\ttfamily\scriptsize move \{object\} to \{target\}}
    };

  \node[outcard, anchor=north west] (invoke)
    at ($(template.south west)+(0,-1.4mm)$) {%
      \textbf{Grounded invoke: with what args}\\[0.4mm]
      {\ttfamily\scriptsize invoke(object-transfer,}\\
      \quad{\ttfamily\scriptsize object="salt",}\\
      \quad{\ttfamily\scriptsize source="counter",}\\
      \quad{\ttfamily\scriptsize target="cabinet")}
    };

  \node[outcard, anchor=north west] (contract)
    at ($(invoke.south west)+(0,-1.4mm)$) {%
      \textbf{Typed signature: what}\\[0.4mm]
      \textbf{$\kappa$}: object-transfer\\
      \textbf{inputs}: object, source, target\\
      \textbf{output}: object at target
    };

  \draw[arrow] (step5.east) -- (outputPanel.west |- step5.east)
    node[label, pos=0.48, above]{pseudocode output};

\end{tikzpicture}

%% file: sections/method.tex
\section{Method}
\label{sec:method}

\subsection{Problem statement}
\label{sec:method:problem}

A \emph{skill library} $\mathcal{L} = \{p_1, \dots, p_n\}$ is a set
of \emph{parent skills}, each a procedural document (markdown
\texttt{SKILL.md}, OpenAPI specification, or MCP server
description). Each parent $p_i$ consists of a sequence of
\emph{procedural units} $u_{i,1}, \dots, u_{i,k_i}$ (markdown
sections, OpenAPI parameter groups, etc.). A
\emph{candidate cluster} $c$ is a set of procedural units, drawn
from one or more parents, hypothesised to describe the same
underlying procedure in different prose.

A \emph{typed contract} $\kappa$ is a structured pseudocode spec
with explicit \texttt{trigger}, \texttt{input\_schema},
\texttt{output\_schema}, \texttt{pre\-conditions},
\texttt{post\-conditions}, and \texttt{side\_effects} fields (full
schema in \autoref{app:ir-schema}). $\kappa$ is the
\emph{structured surrogate} for the prose in $c$: the same
information re-expressed in a form the agent can consume in
one read instead of re-deriving it on every retrieval.

We define the conversion $c \to \kappa$ as \emph{valid} iff $\kappa$
satisfies four deterministic checks (\autoref{sec:method:verifier}):
\textbf{Coverage}, \textbf{Binding}, \textbf{Replacement}, and
\textbf{Risk}. The verifier returns either ``valid, with witness
$\kappa$ and supporting evidence'' or ``invalid, with check-level
rejection reasons''. The verifier produces no soft-classifier
score; any rejection is grounded in one of four named checks.
Given a set $\mathcal{K}$ of validated
contracts, the converted library $\mathcal{L}'$ rewrites each
parent's matching units into \texttt{invoke($\kappa_j$, $\vec{a}$)}
placeholders and adds each $\kappa_j$ as a child skill (reached via
\texttt{invoke}, not direct retrieval; \autoref{sec:method:retrieval}).
Cross-parent factoring (when $|\text{src}(\kappa)| > 1$) is a
secondary benefit; the principal benefit is the
prose-to-pseudocode conversion itself.

\subsection{Pipeline overview}
\label{sec:method:pipeline}

\autoref{fig:pipeline} shows the pipeline: five base stages plus
two LLM-aware post-verifier passes. Parser, candidate proposer,
and refactor (Stages 1, 2, 5) are deterministic; the contract
extractor (Stage 3) issues one \texttt{gpt-4o-mini} call per
candidate cluster; the verifier (Stage 4) is fully deterministic.
The contract IR (full schema in \autoref{app:ir-schema}) is a
typed record with fields \texttt{trigger}, \texttt{input\_schema},
\texttt{output\_schema}, \texttt{pre/postconditions},
\texttt{resources}, \texttt{side\_effects}, \texttt{source\_parents},
and per-call-site \texttt{bindings} filled by BE.

\paragraph{Stage 1: Parser.}
For \texttt{SKILL.md} the parser splits YAML frontmatter from body
and uses markdown headings to define procedural units. For OpenAPI
it groups parameter-handling operations into units. Output:
\texttt{parents.json}, a list of parents each with
\texttt{procedural\_units}, \texttt{scripts}, \texttt{references}.

\paragraph{Stage 2: Candidate proposer.}
For each procedural unit we extract a \emph{frame} tuple
$(\texttt{verb}, \texttt{objects}, \texttt{code\_langs},
\texttt{linked\_scripts})$ via shallow parsing, then embed the unit
text with \texttt{text-embedding-3-small}. Single-linkage clustering
joins units sharing a frame and cosine similarity $\geq 0.65$. The
stage is intentionally \emph{high-recall}; over-clustering is
acceptable because the verifier filters downstream.

\paragraph{Stage 3: Contract extractor (LLM).}
\label{sec:method:stage3-llm}
A single \texttt{gpt-4o-mini} call per candidate cluster produces
a strict-JSON contract draft; the model can refuse via
\texttt{\_extraction\_failed} (structural rejection). Subsequent
rejections are deterministic.

\paragraph{Stage 4: Verifier (4 checks).}
Outputs a profile $\phi(c, \kappa)$ over
\texttt{coverage},
\texttt{binding\_rate},
\texttt{replacement\_rate},
and \texttt{risk}, plus a 3-tier decision (\autoref{sec:method:verifier},
\autoref{sec:method:policy}).

\paragraph{Stage 5: Refactor.}
For each \texttt{auto\_promote} contract, we (i)~detect call-sites
deterministically by walking parents and matching units that
contributed to the cluster, (ii)~apply BE LLM-aware bindings
(below), (iii)~rewrite each parent skeleton by replacing call-site
units with \texttt{invoke($\kappa$, $\vec{a}$)}, then
(iv)~apply RC cleanup. The output is a per-parent
\texttt{*.rewritten.md} for every touched skill plus the index
\texttt{refactored\_library.json}.

\paragraph{Binding Extraction (BE).}
For each deterministic call-site, BE asks \texttt{gpt-4o-mini}
whether the unit is a genuine instance of $\kappa$ and, for each
required input, which substring of the unit text binds it. A
deterministic post-check drops call-sites whose bindings are empty
or whose tokens do not overlap the unit. Passing bindings are
inlined into the rewrite as \texttt{invoke($\kappa$, \{object=
"laptop 1", source="bed 2", dest="desk 1"\})}, grounding the
placeholder per-parent. BE dropped $321/791$ ($41\%$) call-sites as spurious on
\texttt{skills\_500}.

\paragraph{Rewrite Cleanup (RC).}
The deterministic rewrite leaves non-clustered sections (Workflow
lists, Examples, Bundled Resources) verbatim; if any of these
content blocks contradict the child contract (e.g., an Example
using a different action verb), the conflict misleads the agent.
RC is a single-shot verifier-guided rewrite in the spirit of
self-correction \citep{madaan2023selfrefine,gou2024critic}: it asks
\texttt{gpt-4o-mini} to detect and rewrite such conflicts as
\texttt{invoke($\kappa$, $\vec{a}$)} calls, preserving
parent-specific text and introducing no content absent from
either the original parent or the verified contract. A
deterministic post-check ensures every invoked $\kappa$ remains
referenced after cleanup. RC succeeded on $291/292$ ($99.7\%$)
touched parents.

\subsection{The four verifier checks}
\label{sec:method:verifier}

Given a candidate $c$ and contract draft $\kappa$, the verifier
computes four rule-based, check-level scores; unlike learned
step-level verifiers \citep{lightman2024verify,uesato2022process},
the output is a structured rejection profile, not a soft classifier
score. The four checks: \textbf{Coverage} (token recall of
$\kappa$'s trigger/I-O strings against each parent's unit text,
catches misnamed contracts); \textbf{Binding} (for each required
input, fraction of (parent, input) pairs whose unit text overlaps
the input name, catches over-wide clusters); \textbf{Replacement}
(fraction of parents whose unit admits an
\texttt{invoke($\kappa$)}-substitution preserving markdown
structure, catches control-flow entanglement); \textbf{Risk} (AST
scan of contract scripts/resources for unsafe sinks like
\texttt{rm -rf} or undeclared network egress, hard-rejects at
weighted score $\geq 0.80$). Each check catches a distinct,
nearly disjoint class of would-be-promoted candidates; binding is
the dominant first-failed check ($71\%$) by design, because the
deliberately high-recall proposer makes over-wide clusters the most
common failure and binding is their designated filter
(\autoref{app:calibration}, \autoref{tab:reject-breakdown}). Full
threshold definitions and the per-layer rejection bucket are in
\autoref{app:calibration}.

\subsection{Decision policy + calibration (summary)}
\label{sec:method:policy}

The four scores are combined into a scalar promotion score
$s(\phi) = w_b\,\texttt{binding} + w_c\,\texttt{coverage} +
w_r\,\texttt{replacement} - w_s\,\texttt{risk}$ and
mapped to a three-tier decision (\texttt{auto\_promote},
\texttt{review}, \texttt{reject}) by thresholds
$(\tau_{\text{auto}}, \tau_{\text{rev}})$. We calibrate the two
thresholds against $30$ synthetic negative controls from three
single-library classes (same-domain-distinct, near-miss verb-object
permutations, swapped-contract pairings; $n=10$ each), a
benchmark-free counterpart to output calibration
\citep{guo2017calibration} and Chow-style reject options, with $90$
further controls held out to measure the selected point. On
\texttt{skills\_500} a wide range of thresholds achieves $0\%$ FP on
the calibration negatives; for the main result we use the
conservative point
$(\tau_{\text{auto}}, \tau_{\text{rev}})=(0.65, 0.35)$, which
promotes $49$ verified children---the highest-precision operating
point consistent with the verified-refactoring discipline: every
promotion is backed by the strongest check profile the grid admits.
A looser calibrated point $(0.30, 0.10)$ admits $80$ children at
the same $0\%$ FP rate on the calibration negatives and is reported
as a sensitivity variant in \autoref{app:default-policy}.

\subsection{Retrieval-time substitution}
\label{sec:method:retrieval}

\begin{figure}[t]
  \centering
  \resizebox{0.98\columnwidth}{!}{\input{figures/retrieval_bundle.tex}}
  \caption{\textbf{Retrieval-time substitution
  (\S\ref{sec:method:retrieval}).} For each retrieved parent,
  \method{} replaces an \texttt{invoke} placeholder with the content
  the agent needs to act: concrete action templates with bindings,
  the rewritten parent skeleton, and the inlined child contract.}
  \label{fig:retrieval-bundle}
\end{figure}
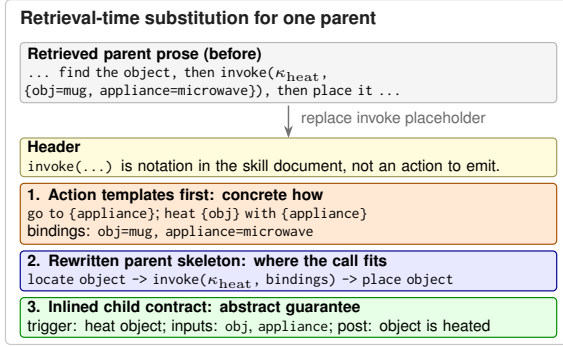

\paragraph{Hierarchical-only retrieval pool.}
At task time the agent retrieves over a pool that contains only
parent skills; promoted child contracts are kept in a separate
internal store. Children are reusable subprocedures designed to be
reached via a parent's \texttt{invoke($\kappa$, $\vec{a}$)}
placeholder, not as standalone top-level capabilities; surfacing
them as standalone results lets the agent read an
under-contextualised child without its wrapping parent
(\autoref{sec:results:retrieval_ablation}).

\paragraph{Substituted skill bundle
  (\autoref{fig:retrieval-bundle}).}
For each retrieved parent whose \texttt{parent\_id} appears in the
refactored library, the retrieval module replaces the parent's
content with a short header (clarifying that \texttt{invoke(...)}
is notation, not an executable verb) followed by three content
blocks: (i)~for every \texttt{invoke($\kappa_j$, $\vec{a}$)} the
verbatim \texttt{original\_unit\_text} (the concrete env-action
sequence the placeholder replaces, e.g.\ \texttt{go to \{recep\}},
\texttt{heat \{obj\} with \{appliance\}}) with extracted bindings;
(ii)~the rewritten parent skeleton; and (iii)~the inlined child
contract specs (trigger, I/O schema, pre/post). This layout puts
the executable action templates in the prefix region the agent
reads first, with the higher-level contract
abstraction available below for cases where the action template
needs to be specialised to a new binding. The substitution adds no
content beyond what the verified pipeline already produced; it only
re-presents it in the order the agent consumes it.

\paragraph{Static binding and its runtime boundary.}
The replacement check certifies that
\texttt{invoke($\kappa$, $\vec{a}$)} can \emph{syntactically}
replace a unit; the bindings $\vec{a}$ are static (index-time).
This suffices when argument slots are environment constants or are
refreshed from a current observation, but cannot supply values that
exist only as episode runtime state: in one failed episode (``cool
a pan''), the agent copies a stale exemplar binding
(\texttt{object="pot 1"}), emits \texttt{cool pan with fridge 1}
before holding the pan, and enumerates receptacles without
re-observing; in a contrasting success the same skill wins because
the agent re-binds from a fresh observation. A runtime expansion
tool (\texttt{sap\_invoke}) would close this gap; we scope it to
future work.

Cross-modality dispatch (the same pipeline on typed OpenAPI libraries
by swapping Stages 1--2) is architecturally supported but not
empirically evaluated here; design in \autoref{app:cross-modality},
typed-API evaluation deferred to future work (\nameref{sec:limitations}).

%% file: figures/retrieval_bundle.tex
\begin{tikzpicture}[
    x=1mm, y=1mm,
    font=\sffamily,
    >={Stealth[length=1.7mm,width=1.2mm]},
    every node/.style={inner sep=0pt},
    panel/.style={
      draw=black!22, fill=white, rounded corners=2pt, line width=0.45pt,
      minimum width=76mm, minimum height=46.6mm,
    },
    title/.style={
      font=\sffamily\scriptsize\bfseries, text=black!88,
      anchor=north west,
    },
    hint/.style={
      font=\sffamily\tiny, text=black!55,
      anchor=north west,
    },
    box/.style={
      draw=black!35, rounded corners=1.5pt, line width=0.45pt,
      text width=70.5mm, align=left,
      font=\sffamily\tiny,
      inner xsep=1mm, inner ysep=0.65mm,
    },
    compactbox/.style={
      box,
      inner ysep=0.45mm,
    },
    arrow/.style={->, draw=black!55, line width=0.55pt},
    label/.style={
      font=\sffamily\tiny, text=black!58, fill=white,
      inner xsep=0.7pt, inner ysep=0.3pt,
    },
  ]

  \node[panel, anchor=north west] (panel) at (0,0) {};

  \node[title] at ($(panel.north west)+(2mm,-1.5mm)$)
    {Retrieval-time substitution for one parent};

  \node[box, anchor=north west, fill=black!4, draw=black!28]
    (raw) at ($(panel.north west)+(2mm,-5.8mm)$)
    {\textbf{Retrieved parent prose (before)}\\
     \texttt{... find the object, then invoke($\kappa_{\mathrm{heat}}$,}\\
     \texttt{\{obj=mug, appliance=microwave\}), then place it ...}};

  \node[label, anchor=west] (substLabel) at ($(raw.south)+(1.4mm,-2.1mm)$)
    {replace invoke placeholder};
  \draw[arrow] ($(raw.south)+(0,-0.3mm)$) -- ($(raw.south)+(0,-4.7mm)$);

  \node[compactbox, anchor=north west, fill=yellow!16, draw=yellow!55!black]
    (header) at ($(panel.north west)+(2mm,-18.7mm)$)
    {\textbf{Header}\\
     \texttt{invoke(...)} is notation in the skill document, not an action to emit.};

  \node[box, anchor=north west, fill=orange!17, draw=orange!65!black]
    (templates) at ($(header.south west)+(0,-0.75mm)$)
    {\textbf{1. Action templates first: concrete how}\\
     \texttt{go to \{appliance\}}; \texttt{heat \{obj\} with \{appliance\}}\\
     bindings: \texttt{obj=mug, appliance=microwave}};

  \node[compactbox, anchor=north west, fill=blue!9, draw=blue!55!black]
    (skeleton) at ($(templates.south west)+(0,-0.75mm)$)
    {\textbf{2. Rewritten parent skeleton: where the call fits}\\
     \texttt{locate object -> invoke($\kappa_{\mathrm{heat}}$, bindings) -> place object}};

  \node[compactbox, anchor=north west, fill=green!10, draw=green!55!black]
    (contract) at ($(skeleton.south west)+(0,-0.75mm)$)
    {\textbf{3. Inlined child contract: abstract guarantee}\\
     trigger: heat object; inputs: \texttt{obj}, \texttt{appliance}; post: object is heated};

\end{tikzpicture}

%% file: sections/experiments.tex
\section{Experimental Setup}
\label{sec:setup}

\subsection{Library and benchmarks}
\label{sec:setup:libraries}

\paragraph{Skill library.}
We use the public Graph-of-Skills \texttt{skills\_500} skillset
\citep{liu2026gos}: 500 markdown skill documents, $5{,}709$ procedural
units after parsing, 54 skills shipping with an executable script.
The skillset covers $\sim$10 domains (37 \texttt{alfworld-*}
embodied-agent skills, 8 \texttt{*-automation} API wrappers,
\texttt{ffmpeg-*}, \texttt{senior-*} software engineering,
mathematical and signal-processing skills, etc.). We selected
\texttt{skills\_500} as the smallest GoS skillset whose content
matches the ALFWorld benchmark domain: \texttt{skills\_200}
(\autoref{app:library-match}) contains zero \texttt{alfworld-*}
skills (verified by \texttt{grep}) and gives the same near-zero
reward across all retrieval modes on ALFWorld.

\paragraph{ALFWorld.}
The unseen split of ALFWorld \citep{shridhar2021alfworld}, 134
text-game household tasks (heat-and-place, cool-and-place,
clean-and-place, pick-and-place, look-at-object, etc.). Each task is
a partially observable environment; success is a binary reward at
the end of an episode of up to $30$ environment-action steps
(skill-retrieval turns do not consume the action budget).

\paragraph{SkillsBench (subset).}
A $10$-task subset of the Docker-environment coding tasks in
SkillsBench \citep{benchflow2026skillsbench}: $6$ easy + $4$ medium
(full task names and snapshot provenance in
\autoref{app:skillsbench-tasks}). Full evaluation was deferred for
budget; sample bias is
discussed in \autoref{sec:results:skillsbench} and
\nameref{sec:limitations}.

\subsection{Models and inference}
\label{sec:setup:models}

\paragraph{ALFWorld agent.} Each game runs up to
$\texttt{max\_steps}=30$ environment actions (the canonical
ALFWorld budget), with \texttt{gpt-4o-mini} at \texttt{temperature=0},
$4$ parallel workers, and a single OpenAI \texttt{seed} shared
across the three retrieval modes within each seed-run.

\paragraph{SkillsBench agent.} \texttt{gpt-5-codex} via the official
\texttt{codex} CLI, routed through an OpenAI-compatible proxy,
inside per-task Docker containers (Harbor harness). Best-of-1 single
attempt per task.

\paragraph{Pipeline LLM.} Contract extraction, BE bindings, and RC
cleanup all call \texttt{gpt-4o-mini} at \texttt{temperature=0}, with
\texttt{text-embedding-3-large} ($d=3072$) for the GoS workspace and
\texttt{text-embedding-3-small} for the proposer.

\subsection{Retrieval modes (baselines + ours)}
\label{sec:setup:modes}

All three modes share the identical agent-facing retrieval interface;
they differ only in library content and visibility budget.
\textsc{all\_full} (Vanilla Skills): full library mounted in context,
agent retrieves by direct file inspection, tokens unbounded.
\textsc{gos}: GoS retrieval \citep{liu2026gos} (embedding seed + graph
rerank) returns top-$K=8$ skills inline. \textsc{sap} (\method{},
ours): the same retrieval algorithm and $K=8$ over the
\emph{refactored} library produced by our pipeline (parent skeletons
+ child contracts as sidecar skills). Every pipeline stage is a CLI
script (\autoref{app:reproducibility}).

\section{Results}
\label{sec:results}

\subsection{Pipeline yield on \texttt{skills\_500}}
\label{sec:results:yield}

The pipeline reduces $5{,}709$ procedural units to $149$ candidate
clusters; at the conservative operating point
$(\tau_{\text{auto}}, \tau_{\text{rev}})=(0.65, 0.35)$ the verifier
auto-promotes $49$ verified child contracts ($0\%$ FP on
calibration controls, $2.2\%$ held out). BE confirms $470/791$ deterministic
call-sites ($59\%$) and drops the rest as spurious; RC successfully
rewrites $291/292$ touched parents ($99.7\%$). Full per-stage yield
is in \autoref{app:pipeline-yield-full}; a looser calibrated point
that admits $80$ children at the same FP rate on the calibration
negatives is reported as a sensitivity variant in
\autoref{app:default-policy}.

\subsection{Main result: ALFWorld 134-game}
\label{sec:results:alfworld}

\begin{figure}[t]
  \centering
  \includegraphics[width=0.95\columnwidth]{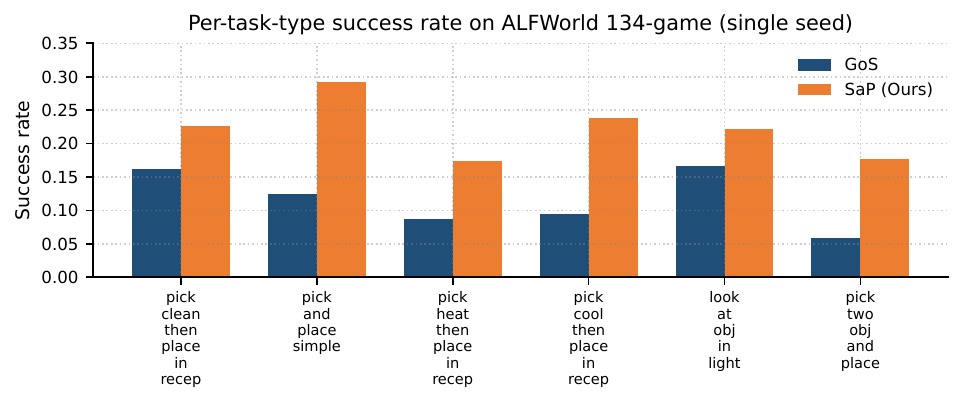}
  \caption{Per-task-type win rate on ALFWorld $134$-game
  (seed $=42$; numerical breakdown in
  \autoref{app:per-task-type}). \method{} beats \textsc{gos} on
  \emph{every} task type, with the largest gains on the multi-step
  heat/cool/place categories.}
  \label{fig:per-task-type}
\end{figure}

\autoref{tab:alfworld-main} reports the main result. On the
ALFWorld unseen split ($134$ games, \texttt{gpt-4o-mini},
$\texttt{temperature}=0$, $\texttt{max\_steps}=30$), pooled across
$3$ seeds ($n = 402$ paired games):

\begin{itemize}[leftmargin=*,topsep=2pt,itemsep=2pt]
\item Reward gain: \method{} wins $82$ vs.\ $47$ for \textsc{gos}
  ($+74\%$ relative; pooled paired McNemar exact
  $p = 8.2 \times 10^{-5}$ on $56$ \method{}-only vs.\ $21$
  \textsc{gos}-only discordant pairs). The single-seed
  $\texttt{seed}{=}42$ point is $30$ vs.\ $16$ wins (McNemar
  $p=0.0043$); the gain is broad---\method{} beats \textsc{gos}
  on \emph{every} task type (\autoref{fig:per-task-type}). Per-seed
  in \autoref{app:multiseed-full}.
\item Token saving \emph{concurrent}: $-22.8 \pm 6.4\%$ input,
  $-17.3 \pm 5.1\%$ output, $-14.5 \pm 4.1\%$ LLM calls per game.
  Reward and tokens co-vary
  (\autoref{sec:analysis:failures}).
\item \method{}'s cross-seed variance ($\pm 4.62$ wins,
  $\pm 15.9$k input tokens) is larger than \textsc{gos}'s
  ($\pm 0.58$, $\pm 0.9$k) because won games close out before the
  $\texttt{max\_steps}{=}30$ budget---on seeds where \method{} wins
  more, it also saves more tokens.
\item Amortization: the one-time ${\sim}\$1.3$ conversion saves
  ${\sim}56$k input tokens per game and pays for itself after
  ${\sim}150$ games at \texttt{gpt-4o-mini} prices, before counting
  the $-14.5\%$ reduction in LLM calls.
\end{itemize}

\input{tables/alfworld_main}

\subsection{SkillsBench: a second-benchmark generality check}
\label{sec:results:skillsbench}

To test whether the representation change extends beyond ALFWorld
we sample $10$ tasks uniformly at random from our SkillsBench
snapshot (task list in \autoref{app:skillsbench-tasks}; full
evaluation deferred for budget, \nameref{sec:limitations}). This
section carries deliberately less evidential weight than ALFWorld:
a secondary generality check via case study and per-task token
analysis, resting on the mechanical input-token reduction rather
than the reward comparison, which is within sampling noise at
$n{=}10$. An exploratory partial run over the remaining tasks is
reported with caveats in \autoref{app:skillsbench-partial}.

\input{tables/skillsbench_main}

\paragraph{Aggregate (\autoref{tab:skillsbench-main}).}
Under \texttt{gpt-5-codex}, \method{} obtains $3/10$ wins versus
$2/10$ for \textsc{gos} at $-43\%$ input tokens.
\textsc{all\_full} (Vanilla) wins $4/10$ at $1.6\times$ the input
tokens; the simultaneous reward-up / token-down effect against
\textsc{gos} replicates the ALFWorld pattern, but does not yet
beat Vanilla in reward at this sample.

\paragraph{Per-task token reductions
  (\autoref{tab:skillsbench-per-task}).}
\method{} reduces input tokens on $8/10$ tasks (range $-14\%$ to
$-82\%$); the two tasks where it uses more are both joint wins
where \method{} wrote more code to reach the same answer. Two
\method{}-run errors (\texttt{court-form-filling},
\texttt{data-to-d3}) failed mid-execution
(\texttt{NonZeroAgentExitCode}, counted as $0$ reward); their
input tokens were already $-59\%$ / $-15\%$ vs.\ \textsc{gos}
when the crash occurred, so retrieval was not the bottleneck here.

\paragraph{Case study: \texttt{citation-check}
  (\method{}-only win).}
\texttt{gpt-5-codex} imposes an $8$K shell-output cap. \textsc{gos}'s
raw \texttt{citation-management/SKILL.md} is $8{,}354$ tokens; the
agent reads a body cut mid script-listing, invokes the bundled
\texttt{validate\_citations.py} whose output schema mismatches the
task's required \texttt{answer.json}, attempts a from-scratch
reimplementation, and runs out of budget. \method{}'s refactored
version of the same skill is a $3{,}566$-token rewritten skeleton
plus child contract, fitting under the cap; the agent reads it
whole, treats the script as reference only, writes its own
\texttt{bibtexparser}-based extraction, and produces the correct
output. The SkillsBench mechanism is the content-density analogue
of ALFWorld's per-retrieval-density: prose forces the agent to
recover lost content, and recovery does not always fit the budget.

\subsection{Pipeline component ablation}
\label{sec:results:ablation}

We add each LLM-aware pass incrementally on a fixed $20$-game
ALFWorld subset (idx $0$--$19$): deterministic-only $\to$ +RC
$\to$ +BE+RC = $0.05 \to 0.10 \to 0.15$ ($+5$\,pp per step),
matching \textsc{gos} on the same subset and exceeding it on the
full $134$ (\autoref{tab:alfworld-main}; full table in
\autoref{tab:component-ablation}). BE-alone (without RC) was not
run.

\textbf{Each pass repairs a specific failure mode.} RC fixes stale
syntax in unrewritten content: e.g., a preserved \texttt{Example}
block still emits \texttt{Action: put laptop 1 in/on desk 1}, a
verb the ALFWorld engine does not parse (it accepts only
\texttt{move X to Y}); RC detects the contradiction against the
child contract and rewrites the Example so the bundle propagates
the engine-accepted verb. BE drops $321/791$ deterministic
call-sites ($41\%$) as spurious---units sharing keywords with the
contract but describing a distinct procedure---which the per-unit
token filter retains but a prompt-aware judgment removes. Full
pre/post diff and the spurious-call-site catalogue are in
\autoref{app:phase55-diff} and \autoref{app:be-spurious}.

\subsection{Cross-model-family pipeline validation}
\label{sec:results:crossmodel}

Since extraction, BE, and RC depend on instruction following and
structured output, we rerun all three with a non-OpenAI
family---\texttt{deepseek-v4-pro}, a reasoning-style
model---keeping verifier, thresholds, policy, and embeddings
identical (\autoref{tab:crossmodel-pipeline}). One adaptation was
required, and it locates the porting cost: completion budgets sized
for a non-reasoning model truncate a reasoning model's outputs
(hidden deliberation consumes the budget). Every truncated output
was caught by the gates---parsed as failure, never promoted---and
raising the budgets (a configuration change; no prompt or verifier
changes) resolved it. Under identical gates, contract-level
promotion is nearly model-independent ($44$ vs.\ $49$ children);
the binding pass diverges most ($16\%$ vs.\ $59\%$ call-sites
confirmed). Divergence manifests as refusals and conservative
yield, never as unfaithful contracts silently entering the library.

\input{tables/crossmodel_pipeline}

\subsection{Retrieval-pool ablation: hierarchical separation}
\label{sec:results:retrieval_ablation}

\method{} separates top-level parent retrieval from the internal
child-contract store; children are reached only via parent
\texttt{invoke} placeholders. Ablating this hierarchy by allowing
child contracts to surface as top-level results drops reward
$22.4\% \to 16.4\%$ on \texttt{seed}=42 ($-27\%$; $30 \to 22$
wins): children appear in the top-$K$ in $\sim 8\%$ of games, and
the agent then reads the contract spec without the parent's action
template, breaking the contract+template pair.

\subsection{Bundle-component ablation}
\label{sec:results:bundle_ablation}

The substituted bundle changes several things at once: it exposes a
typed contract, surfaces action templates first, and is shorter
than raw prose. \autoref{tab:bundle-ablation} disentangles these
under the identical protocol and retrieval as the main result (only
the content served after retrieval differs). Three findings.
(i)~\emph{The action template is necessary}: removing it collapses
reward below even the prose baseline ($13$ vs.\ template-only's
$25$, paired McNemar $p=0.043$)---a typed signature without
environment-level syntax cannot be acted on, mirroring
\autoref{sec:results:retrieval_ablation} from the other side.
(ii)~\emph{The gain is not a length effect}: prose truncated to the
identical character budget recovers almost none of the gap ($18$
vs.\ $16$), remaining $12$ wins below the full bundle.
(iii)~\emph{The signals are complementary}: template-only sits
between contract-only and the full bundle. We report the one
significant pairwise test and present the ordering descriptively;
single-seed $n{=}134$ has limited power for smaller contrasts.

\input{tables/bundle_ablation}

\subsection{Verifier threshold calibration}
\label{sec:results:calibration}

We grid-search $(\tau_{\text{auto}}, \tau_{\text{rev}})$ against
$30$ synthetic negative controls drawn from three single-library
classes: \texttt{B\_same\_domain\_distinct} (a contract from a
different sub-domain), \texttt{C\_near\_miss} (verb-object
permutations of a real contract), \texttt{D\_swapped\_contract} (a
real contract assigned to the wrong cluster); $n=10$ each. On
\texttt{skills\_500} a wide region of the grid achieves $0\%$ FP
on these negatives. We use the conservative point
$(\tau_{\text{auto}}, \tau_{\text{rev}}) = (0.65, 0.35)$, promoting
$49$ verified children---including genuinely recurring
sub-procedures (\texttt{alfworld-navigation},
\texttt{appliance-state-modifier}) whose extraction is the central
thesis and whose contracts the agent demonstrably reads and acts on
(\autoref{sec:analysis:failures}). A looser calibrated point
$(0.30, 0.10)$ admits $80$ at the same $0\%$ FP on the
calibration negatives; its additional promotions carry weaker check profiles,
so we report it as a library-level sensitivity variant
(\autoref{app:default-policy}) without downstream evaluation. Full
grid and per-layer breakdown in \autoref{app:calibration}.

%% file: tables/alfworld_main.tex
\begin{table*}[t]
\centering
\small
\setlength{\tabcolsep}{6pt}
\begin{tabular}{lcrrrr}
\toprule
Mode & $n$ games & wins & in / g (k) & out / g & calls / g \\
\midrule
\textsc{all\_full}$^{\dagger}$ & $20$ & $0$ & $833.8$ & --- & --- \\
\textsc{gos}                   & $134 \times 3$ & $47$\;\scriptsize($11.7\%$) & $247.8 \pm 0.9$ & $1{,}935 \pm 29$ & $39.7 \pm 0.5$ \\
\textbf{\textsc{sap} (\method{}, ours)}
                               & $\mathbf{134 \times 3}$ & $\mathbf{82}$\;\scriptsize($\mathbf{20.4\%}$) & $\mathbf{191.4 \pm 15.9}$ & $\mathbf{1{,}599 \pm 82}$ & $\mathbf{33.9 \pm 1.5}$ \\
\multicolumn{2}{r}{$\Delta$ \method{} vs.\ \textsc{gos}:}
                               & \scriptsize\textbf{$p\!=\!8.2\!\times\!10^{-5}$}
                               & \scriptsize($-22.8 \pm 6.4\%$)
                               & \scriptsize($-17.3 \pm 5.1\%$)
                               & \scriptsize($-14.5 \pm 4.1\%$) \\
\bottomrule
\end{tabular}
\caption{Main result on ALFWorld unseen split, \texttt{gpt-4o-mini},
\texttt{temperature=0}, $\texttt{max\_steps}=30$. \textsc{gos} and
\method{} numbers are pooled across $3$ seeds ($n = 402$ paired
games; $\pm$ values are cross-seed std on per-game means). \method{}
simultaneously raises reward \emph{and} saves tokens on every seed
(paired McNemar exact $p = 8.2 \times 10^{-5}$ on $56$ \method{}-only
vs.\ $21$ \textsc{gos}-only discordant pairs; single-seed
$\texttt{seed}{=}42$ McNemar $p = 0.0043$).
$^{\dagger}$\textsc{all\_full} is a $20$-game pilot on
$\texttt{seed}{=}42$ only ($\sim 3.4\times$ \textsc{gos}'s input tokens,
$0$ wins; full-library prompts overwhelm \texttt{gpt-4o-mini} and
the per-game wallclock is dominated by rate-limit retries).
Per-seed breakdown in \autoref{tab:multiseed-full}.}
\label{tab:alfworld-main}
\end{table*}

%% file: tables/skillsbench_main.tex
\begin{table}[t]
\centering
\small
\setlength{\tabcolsep}{4pt}
\begin{tabular}{lrr}
\toprule
Mode & Avg R & Input tok \\
\midrule
\textsc{all\_full} & 0.40 (4/10) & 30.80 M \\
\textsc{gos} & 0.20 (2/10) & 33.94 M \\
\textbf{\textsc{sap} (ours)} & \textbf{0.30 (3/10)} & \textbf{19.21 M} ($-$43\%) \\
\bottomrule
\end{tabular}
\caption{SkillsBench $10$-task subset, \texttt{gpt-5-codex}.
\method{} obtains $+1$ win over \textsc{gos} at $-43\%$ input
tokens; \textsc{all\_full} wins one more but at $1.6\times$ the
input tokens. Subset is sample-biased ($6$ easy + $4$ medium);
full evaluation deferred for budget.}
\label{tab:skillsbench-main}
\end{table}

%% file: tables/crossmodel_pipeline.tex
\begin{table}[t]
  \centering
  \small
  \setlength{\tabcolsep}{3pt}
  \begin{tabular}{lcc}
    \toprule
    \textbf{Stage metric} & \shortstack{\texttt{gpt-4o-}\\\texttt{mini}}
      & \shortstack{\texttt{deepseek-}\\\texttt{v4-pro}} \\
    \midrule
    Contract drafts attempted & 149 & 149 \\
    Extractor refusals        & 18  & 54 \\
    Malformed drafts          & 0   & 0 \\
    Children promoted         & 49  & 44 \\
    BE call-sites kept        & 470/791 & 291/1{,}813 \\
    RC cleanups accepted      & 291/292 & 223/229 \\
    \bottomrule
  \end{tabular}
  \caption{\textbf{Cross-model-family pipeline validation.} The
  three LLM-aware stages rerun with \texttt{deepseek-v4-pro} under
  the identical deterministic verifier, thresholds, and promotion
  policy. Model-family divergence is absorbed as explicit refusals
  and post-check rejections---never as corrupted contracts entering
  the library.}
  \label{tab:crossmodel-pipeline}
\end{table}

%% file: tables/bundle_ablation.tex
\begin{table}[t]
  \centering
  \small
  \setlength{\tabcolsep}{3pt}
  \begin{tabular}{lc}
    \toprule
    \textbf{Bundle content served after retrieval} & \textbf{wins/134} \\
    \midrule
    Full bundle (templates + skeleton + contracts) & \textbf{30} \\
    Template-only (contract block removed)         & 25 \\
    Length-matched raw prose (same char budget)    & 18 \\
    \textsc{gos} raw prose, unbounded              & 16 \\
    Contract-only (template block removed)         & 13 \\
    \bottomrule
  \end{tabular}
  \caption{\textbf{Bundle-component ablation}, ALFWorld
  $134$-game, seed $=42$, identical protocol and retrieval as
  \autoref{tab:alfworld-main}. Serving the typed contract without
  the concrete action template falls below even the prose baseline
  (template-only vs.\ contract-only paired McNemar $p=0.043$);
  length-matched prose recovers almost none of the gap.}
  \label{tab:bundle-ablation}
\end{table}

%% file: sections/analysis.tex
\section{Analysis}
\label{sec:analysis}

\subsection{Mechanism: structure breaks the retrieval-action loop}
\label{sec:analysis:failures}

\method{}'s bundle delivers two structured signals (typed contract
+ inlined action template) in one response, while \textsc{gos}
forces the agent to locate both pieces inside a long mixed prose
body---an instance of the ``lost-in-the-middle'' degradation
\citep{liu2024lostmiddle}. Three observations support this:
post-retrieval correctness (\autoref{tab:post-retrieval}), a
per-game loop signature (\autoref{app:mechanism-split}), and a
trace-level case study.

\paragraph{Post-retrieval action correctness.}
\method{}'s next env action after a retrieval succeeds $29.3\%$
of the time vs.\ $20.5\%$ for \textsc{gos} (pooled $3$ seeds), with
the gap widening to $29.8\%$ vs.\ $19.0\%$ ($1.57\times$) on
\texttt{READ\_SKILL} events---when the agent explicitly looks up a
skill, \method{}'s bundle enables a correct immediate action much
more reliably, before any second-look loop begins.

\input{tables/post_retrieval_action}

\paragraph{Loop is the unit of failure.}
Aggregated to game scale (\autoref{app:mechanism-split}): lost games
issue $\sim 4\times$ more \texttt{SkillRequest}s and
``\texttt{Nothing happens.}'' than won games \emph{within either
mode}; \method{} cuts \texttt{SkillRequest}/game by $\sim 30\%$ on
lost games. The retrieval-pool ablation
(\autoref{sec:results:retrieval_ablation}) confirms the dependency
on \emph{both} signals: when promoted contracts are surfaced as
standalone results without their wrapping parent, reward drops $27\%$.

\paragraph{Exposure and retrieval shift.}
$62.9\%$ of the $402$ pooled games read at least one substituted
bundle (median $1$ distinct promoted contract, max $3$); bundles
also arrive via failure-triggered injections, so coverage is an
exposure statistic, not a win predictor. Rewriting also shifts
embedding retrieval, but locally: top-$1$ agreement with
\textsc{gos} is $62.7\%$ per seed, and \textsc{gos}'s top skill
stays within \method{}'s top-$5$ in $93$--$96\%$ of tasks
(\autoref{app:retrieval-rank}). The contrasts in
\autoref{sec:results:bundle_ablation} hold retrieval fixed and are
free of this shift; the headline comparison includes both effects.

\paragraph{Trace contrast.}
On \texttt{idx\_36} (``put a hot mug in a cabinet'') \textsc{gos}
wins in $22$ steps and $4$ \texttt{SkillRequest}s after three
near-syntax errors; \method{} wins in $7$ steps with one
\texttt{READ\_SKILL}, emitting the bundle's templates in order
without ``\texttt{Nothing happens.}'' Full traces:
\autoref{app:case-study-1}, \autoref{app:case-study-2}.

\subsection{What BE and RC each repair}
\label{sec:analysis:cleanup}

The $+5$\,pp/$+5$\,pp component ablation
(\autoref{tab:component-ablation}) isolates two failure modes
already detailed in \autoref{sec:results:ablation}: \textbf{RC}
fixes parent-residual verb conflicts (e.g., an \texttt{Example}
block emitting \texttt{put X in/on Y} where the engine accepts only
\texttt{move X to Y}); \textbf{BE} drops $41\%$ of deterministic
call-sites as spurious. Both passes target distinct upstream causes
of the same prose-induced loop: stale syntax in unrewritten content
(RC) and false-positive cluster membership (BE).

%% file: tables/post_retrieval_action.tex
\begin{table}[t]
\centering
\small
\setlength{\tabcolsep}{3pt}
\begin{tabular}{lrrrr}
\toprule
Mode & events & OK\,\% & ``Noth.''\,\% & re-retr.\,\% \\
\midrule
\textsc{gos}          & $2{,}531$ & $20.5$ & $54.8$ & $24.7$ \\
\textbf{\textsc{sap}} & $\mathbf{1{,}757}$ & $\mathbf{29.3}$ & $\mathbf{43.7}$ & $27.1$ \\
\midrule
\multicolumn{5}{l}{\footnotesize \texttt{READ\_SKILL} events only:} \\
\textsc{gos}          & $2{,}129$ & $19.0$ & $61.4$ & $19.5$ \\
\textbf{\textsc{sap}} & $\mathbf{1{,}355}$ & $\mathbf{29.8}$ & $\mathbf{54.2}$ & $15.9$ \\
\bottomrule
\end{tabular}
\caption{Post-retrieval next-action outcome (pooled $3$ seeds;
per-game split in \autoref{app:mechanism-split}). The three columns
partition all retrieval events: OK\,\% = emitted an accepted env
action; ``Noth.''\,\% = emitted an action returning
``\texttt{Nothing happens.}''; re-retr.\,\% = issued another
\texttt{SkillRequest} without acting. \method{} is $1.43\times$
($1.57\times$ on \texttt{READ\_SKILL}) more likely to act
successfully and issues $31\%$ fewer retrievals.}
\label{tab:post-retrieval}
\end{table}

%% file: sections/conclusion.tex
\section{Conclusion}
\label{sec:conclusion}

Our results raise a broader question: should skill libraries be
authored in pseudocode directly? Markdown optimises for human
authors, but LLM agents pay the prose-to-structure conversion cost
\emph{on every retrieval}. \method{} shows this cost is recoverable
post hoc with deterministic quality control---paid at index time
rather than at every read---for markdown skill libraries, the
setting we evaluate. If the same holds for typed catalogues
(OpenAPI, MCP; designed but not evaluated,
\autoref{app:cross-modality}), publishers could pre-substitute at
index time while domain experts keep writing prose. More broadly,
\method{}'s discipline---each LLM extraction is a hypothesis
checked by deterministic rules---transfers to other
prose-to-structure tasks; the rule generalises: accept with a
witness, reject with a named cause.

%% file: sections/limitations.tex
\section*{Limitations}
\label{sec:limitations}

We list the three limitations most likely to affect a reader's
interpretation of the numbers in \autoref{sec:results}; secondary
limitations (single agent-model sweep, synthetic-only negative
controls, no human user study, English-only, deferred recursive
split / repair loops) are detailed in
\autoref{app:limitations-extended}. The single-LLM-family pipeline
limitation of the original submission is now partially addressed by
the cross-model validation in \autoref{sec:results:crossmodel},
which covers the pipeline side but not the agent side.

\paragraph{Benchmark sample on SkillsBench.}
SkillsBench results are reported on $10$ tasks sampled uniformly at
random from our SkillsBench snapshot
(\autoref{app:skillsbench-tasks}). We attempted a full-scale
evaluation but ended it early due to per-task LLM budget and
infrastructure errors; we do not include that partial run in the
main body to avoid drawing a conclusion from an incomplete,
non-random sample, and instead report it with caveats as
exploratory material in \autoref{app:skillsbench-partial}.
On the $10$ tasks reported, \method{} beats \textsc{gos} in both
reward ($3$ vs.\ $2$ wins) and input tokens ($-43\%$);
\textsc{all\_full} (Vanilla) wins one more than \method{} but at
$1.6\times$ the input tokens, so the simultaneous-gain claim against
\textsc{gos} replicates while
the claim against \textsc{all\_full} does not yet at this sample
size. Our SkillsBench claim is therefore narrower than ALFWorld:
the $10$-task sample is intended as a generality check via case
study and per-task token analysis
(\autoref{sec:results:skillsbench}, \autoref{app:skillsbench-per-task}),
not a tournament-scale comparison; the per-task token reduction
holds on $8/10$ tasks regardless of binary task reward, indicating
the representation effect is consistent at task granularity.

\paragraph{Library scope: one library, markdown only.}
The pipeline ran end-to-end only on the \texttt{skills\_500} GoS
library. Larger GoS skillsets (\texttt{skills\_1000},
\texttt{skills\_2000}; \citealt{liu2026gos}) were not re-run.
Typed-API libraries (Stripe OpenAPI 587 endpoints, GitHub OpenAPI
1153 endpoints) were prepared at the parser+candidate stage in earlier
work but \emph{not} main-experimented under the current
verified-refactoring pipeline; the cross-modality dispatch claims in
\autoref{app:cross-modality} are therefore architectural, not
empirical, in this submission.

\paragraph{Replacement is static, not run-time.}
The verifier's \emph{replacement} check tests whether
\texttt{invoke($\kappa$, $\vec{a}$)} can syntactically replace the
unit, not whether the agent at run time correctly grounds the call;
\autoref{sec:method:retrieval} characterises the boundary with a
worked failure/success pair (static bindings suffice when argument
slots are environment constants or refreshed from observations, and
fail when slots are episode-specific runtime state). A run-time
mechanism that explicitly expands the call (e.g., a
\texttt{sap\_invoke} tool that resolves $\vec{a}$ against current
state) would close this gap; we did not implement it because the
GoS retrieval interface already lets the agent read child contracts
on demand. Stronger claims about ``hierarchical invocation'' would
require the run-time mechanism.

%% file: sections/ethics.tex
\section*{Ethics Statement}
\label{sec:ethics}

\paragraph{Data provenance.}
All skill libraries used are publicly available: the Graph-of-Skills
\texttt{skills\_500} skillset is published on HuggingFace
\citep{liu2026gos}; ALFWorld \citep{shridhar2021alfworld} is an
open research benchmark; SkillsBench \citep{benchflow2026skillsbench}
is open-sourced on GitHub. No personally identifiable information,
proprietary data, or human subjects are involved.

\paragraph{Model use.}
We use \texttt{gpt-4o-mini} for the pipeline LLM and ALFWorld agent,
\texttt{gpt-5-codex} for the SkillsBench agent (routed through an
OpenAI-compatible proxy), and \texttt{deepseek-v4-pro} for the
cross-model pipeline validation
(\autoref{sec:results:crossmodel}). Total LLM expenditure for the
experiments reported in this paper is approximately \$80 over the
course of development, including the camera-ready ablations.

\paragraph{Safety: the risk verifier check.}
The fourth verifier check (\autoref{sec:method:verifier}) scans
candidate child contracts for unsafe sinks
(\texttt{subprocess.shell}, \texttt{eval}, file deletion, undeclared
network egress) and rejects clusters that introduce them.
This is intentionally conservative: a recurring snippet that legitimately
needs \texttt{rm} or \texttt{requests.post} must declare it in
\texttt{side\_effects} to pass. We do not claim the risk check is a
complete safety solution---it does not detect prompt injection,
adversarial chains, or higher-level reasoning failures---but it does
prevent the most direct failure mode where library factorization
silently aggregates unsafe operations.

\paragraph{Dual use.}
A malicious actor could in principle apply the same pipeline to
compress a library of harmful tools. The method is content-neutral:
\method{} factors whatever subprocedures recur, without judging
whether the underlying skills are beneficial. We discourage such
applications and do not provide a harm-classification step.

\paragraph{Deployment governance.}
Automatic refactoring may simplify or relocate caveats that matter
in real workflows. Before using \method{}-refactored contracts in
compliance-, safety-, or access-control-critical settings, we
recommend human review of promoted contracts, audit logging of the
pipeline's accept/reject decisions (which the verifier's named
rejection causes support directly), and task-level validation.
The risk check screens direct unsafe operations but does not
guarantee preservation of higher-level policy constraints or
task-specific warnings.

\paragraph{Reproducibility.}
We release the full pipeline as a sequence of CLI scripts
(\autoref{app:reproducibility}), the refactored
\texttt{skills\_500} library used in experiments, all 134-game and
10-task evaluation outputs, and the calibrated policy file. ALFWorld
and SkillsBench setup instructions are documented in the appendix.
At \texttt{temperature=0} the runs are nominally deterministic but
in practice modulo low-level non-determinism in the OpenAI API; we
flag this in \nameref{sec:limitations}.

\paragraph{Compute and environmental impact.}
End-to-end pipeline (parser through RC) takes $\sim$30
minutes wallclock and $\$1.3$ LLM cost on \texttt{skills\_500}.
ALFWorld 134-game evaluation per mode is $\sim$1.5 hours wallclock
and $\$3$--$5$ at \texttt{gpt-4o-mini} prices.

\paragraph{AI assistance.}
We used AI coding assistants to help check code correctness and
to proofread paper grammar. All claims, results, and analyses were
verified by the authors.

%% file: sections/acknowledgments.tex
\section*{Acknowledgments}

This work is supported by OPPO.

%% file: sections/appendix.tex
\section{Library/benchmark match note}
\label{app:library-match}

A pre-experiment sanity check on \texttt{skills\_200} (the GoS
runner default) returned identical $5\%$ reward across all four
modes on ALFWorld. Inspection showed \texttt{file-organizer} and
\texttt{sqlite-map-parser} surfacing for the query ``find tomato in
cabinets''; \texttt{skills\_200} has \emph{zero}
\texttt{alfworld-*} skills, so the agent's behaviour is
indistinguishable from \textsc{none}. Switching to
\texttt{skills\_500} ($37$ \texttt{alfworld-*} skills) immediately
makes retrieval contentful. Cautionary methodological note: runner
defaults do not guarantee library/benchmark domain match, and a
30-second \texttt{grep} is sufficient to detect mismatch before
consuming $80$ game-hours of benchmark.

\section{Pipeline component ablation table}
\label{app:component-ablation-table}

\input{tables/component_ablation}

\autoref{tab:component-ablation} accompanies
\autoref{sec:results:ablation}: each LLM-aware pass added
incrementally on the fixed $20$-game pilot subset.

\section{RC cleanup diff (full)}
\label{app:phase55-diff}

Pre-cleanup Example block in \texttt{alfworld-object-transporter}:
\begin{quote}\small\ttfamily
Action: put laptop 1 in/on desk 1\\
Observation: You put the laptop 1 in/on the desk 1.
\end{quote}
Post-cleanup:
\begin{quote}\scriptsize\ttfamily
Action: invoke(\\
\quad alfworld-object-management,\\
\quad \{object="laptop 1",\\
\quad ~~source\_receptacle="bed 2",\\
\quad ~~target\_receptacle="desk 1"\})
\end{quote}
The verb mismatch (\texttt{put} vs.\ \texttt{move}) is rectified
because the child contract's \texttt{output\_schema} uses
\texttt{move}, which RC propagates back into the Example.

\section{SkillsBench per-task breakdown}
\label{app:skillsbench-per-task}

\input{tables/skillsbench_per_task}

\section{SkillsBench \texttt{citation-check} mechanism}
\label{app:skillsbench-citation}

\texttt{citation-check} is a \method{}-only win
(\autoref{tab:skillsbench-main}). Mechanism: under \texttt{gpt-5-codex}
the runtime imposes an $8$K-token shell-output cap; the \textsc{gos}
raw \texttt{citation-management/SKILL.md} is $8{,}354$ tokens, so the
agent reads a body cut mid script-listing. It first tries the bundled
\texttt{validate\_citations.py}, finds the output format does not
match the task's required \texttt{answer.json} schema, then attempts
to re-implement the validator from scratch and exhausts its compute
budget. \method{}'s refactored version of the same skill is a
$3{,}566$-token rewritten skeleton plus child-contract block, which
fits the cap; the agent reads the full skeleton, treats the bundled
script as reference only, immediately writes its own
\texttt{bibtexparser}-based extraction, and produces the correct
\texttt{answer.json}. The mechanism is the SkillsBench analogue of
the ALFWorld retrieval-action loop: in both cases prose's lower
information density forces the agent to recover lost content, and in
both cases that recovery does not always succeed within the budget.

\section{SkillsBench subset task list}
\label{app:skillsbench-tasks}

The $10$-task subset comprises $6$ easy tasks
(\texttt{court-form-filling}, \texttt{dialogue-parser},
\texttt{fix-build-agentops}, \texttt{fix-build-google-auto},
\texttt{offer-letter-generator},
\texttt{powerlifting-coef-calc}) and $4$ medium tasks
(\texttt{citation-check}, \texttt{data-to-d3}, \texttt{edit-pdf},
\texttt{exceltable-in-ppt}).

\paragraph{Snapshot provenance.}
SkillsBench is a living inventory: \citet{benchflow2026skillsbench}
describe it as the benchmark's \emph{current} contents, and the count
has moved across releases. The snapshot we obtained contains $94$
task directories; the accompanying repository README states $87$
coding tasks, and we could not establish which $87$ of the $94$ the
published figure refers to. We therefore quote no denominator for our
own sample and report the subset by name above. The $84$ directories
outside the subset are the set used for the exploratory partial run
(\autoref{app:skillsbench-partial}). Any comparison to published
SkillsBench numbers should be read against the release those numbers
were computed on, not against this snapshot.

\section{Exploratory SkillsBench partial run}
\label{app:skillsbench-partial}

We attempted to extend the \method{} evaluation to the remaining
SkillsBench tasks but ended the run early on per-task LLM budget;
$42$ task attempts were recorded before the stop. Of these, $21$
completed cleanly and \method{} passes $6$ ($28.6\%$); the other
$21$ attempts failed on infrastructure errors (proxy websocket and
DNS failures, job interruptions) rather than on agent behavior, and
one agent-side timeout. We report this run as \emph{exploratory
only}, for three reasons stated up front: the task sample is a
prefix of the remaining tasks, not a random draw; no paired
\textsc{gos} or \textsc{all\_full} baseline was run on these tasks;
and the infrastructure failure rate ($50\%$ of attempts) makes the
completed subset self-selected. Within those constraints, the clean
completion rate is consistent in magnitude with the published
GoS-paper numbers on the full benchmark
(\autoref{app:vs-published}), and no completed task contradicts the
input-token reductions observed on the $10$-task subset.

\section{Contract IR field schema}
\label{app:ir-schema}

The full IR field schema for a child contract $\kappa$:

\begin{itemize}[leftmargin=*,topsep=2pt,itemsep=2pt]
\item \texttt{id}, \texttt{trigger}: stable identifier and one-line
  natural-language invocation condition.
\item \texttt{input\_schema}, \texttt{output\_schema}: typed
  required / optional inputs and outputs.
\item \texttt{preconditions}, \texttt{postconditions}: state
  predicates before and after invocation.
\item \texttt{resources}: bundled scripts, references, external
  tools.
\item \texttt{side\_effects}: filesystem, network, user-visible
  output.
\item \texttt{source\_parents}: which parent skills the cluster was
  drawn from.
\item \texttt{bindings}: per-call-site map from parent unit text to
  $\kappa$'s input fields (filled by BE).
\item \texttt{verifier\_stats}: the four check scores.
\end{itemize}

A worked example contract (\texttt{alfworld-object-management}) and
its OpenAPI counterpart (a Stripe paginated-list contract from the
smoke-test corpus) are in the released code release.

\section{Extended limitations}
\label{app:limitations-extended}

The main-text limitations (\nameref{sec:limitations}) cover
single-seed, SkillsBench subset, single library, and static
replacement. We list secondary limitations here for completeness.

\paragraph{Single agent / single model.}
ALFWorld evaluations use \texttt{gpt-4o-mini}; SkillsBench uses
\texttt{gpt-5-codex} (best-of-1, no retries). Behavior under
substantially weaker (e.g., $7$B class) or stronger (e.g.,
reasoning-heavy) models is not tested. The $+50\%$ relative gain may
be amplified or attenuated under different agent capabilities; a
sweep across model scales is needed.

\paragraph{Pipeline LLM family coverage.}
The cross-model validation (\autoref{sec:results:crossmodel})
covers extraction, BE, and RC with one non-OpenAI family
(\texttt{deepseek-v4-pro}); further families (Claude, Gemini) and,
more importantly, whether the downstream $+5$pp / $+5$pp ablation
effects reproduce with a non-OpenAI \emph{agent}, remain untested.
Per-pass pipeline cost is under $\sim\$5$ per family, so broader
sweeps are feasible.

\paragraph{Negative controls are synthetic.}
Calibration uses $30$ controls of three synthetic classes
(\texttt{B\_same\_domain\_distinct}, \texttt{C\_near\_miss},
\texttt{D\_swapped\_contract}), with $90$ more held out
(\autoref{app:calibration}). Real adversarial inputs---e.g., a
candidate cluster that looks coherent to the LLM but is semantically
broken---are not represented. Both the $0\%$ FP on the calibration
draw and the $2.2\%$ on the held-out draw are therefore
\emph{lower bounds} on practical FP; field FP on real mis-clustered
candidates would require human gold labels we do not have.

\paragraph{No human user study.}
We use deterministic proxies (binary task reward, token count,
verifier-call counts). We do not measure downstream user-perceived
agent quality, latency under realistic load, or developer-side
maintenance cost reduction. The latter is arguably the most direct
motivation for library factorization but is not in scope here.

\paragraph{English-only.}
All parsed skills are in English. Cross-lingual \texttt{SKILL.md} or
non-English OpenAPI documentation is untested.

\paragraph{Recursive split and repair loops deferred.}
The pipeline includes scripts for (a)~recursive splitting of
over-wide rejected clusters into sub-clusters that may pass
verification, and (b)~a $1$--$2$-round LLM repair loop on borderline
rejects. Both are implemented but disabled by default in this
paper's results, to keep the ablation cleanly attributable to the
two LLM-aware passes we evaluate. Enabling either is expected to
further increase promotion yield.

\section{Looser-calibration variant}
\label{app:default-policy}

The main results in \autoref{sec:results:alfworld} use the
conservative operating point $(\tau_{\text{auto}},
\tau_{\text{rev}})=(0.65,0.35)$, which promotes $49$ verified child
contracts at $0\%$ FP on the calibration negatives ($2.2\%$ held
out). The calibration grid
(\autoref{tab:calibration}) shows that a looser calibrated point
$(\tau_{\text{auto}}, \tau_{\text{rev}})=(0.30,0.10)$ also achieves
$0\%$ FP on the calibration negatives and admits $80$
children---$31$ more than the main point. We keep the conservative
$49$-child point as the headline because every one of its
promotions is backed by the strongest check profile the grid
admits, matching the verified-refactoring discipline the paper
argues for; the additional $31$ candidates admitted at the looser
point pass on weaker binding and coverage margins. We report the
$80$-child point here as a library-level sensitivity observation
(promotion counts and check profiles); we did not evaluate it
downstream under the final pipeline, and note that synthetic-control
FP is a lower bound on practical FP
(\autoref{app:limitations-extended}), so looser thresholds carry
correspondingly more semantic risk per promotion.

\section{Calibration grid and verifier-layer breakdown}
\label{app:calibration}

\autoref{tab:calibration} reports the full threshold grid on
\texttt{skills\_500}. Real candidates: $n=149$; calibration
negatives: $n=30$ (classes \texttt{B}, \texttt{C}, \texttt{D}; $10$
each---class \texttt{A\_cross\_domain} requires multiple libraries
and is not available for the single-library run). Promotion count is
monotone non-decreasing as thresholds drop because no row in the
feasible region violates FP $\leq 5\%$.

\paragraph{Held-out controls.}
Thresholds selected on $30$ controls would be fit to that draw, so we
measure the chosen point against a held-out set of $90$ further
controls ($30$ per class) from the same generator under a different
seed. At $(0.65, 0.35)$, $81/90$ are rejected, $7$ are routed to
\texttt{review}, and $2$ are auto-promoted---a held-out
false-positive rate of $2.2\%$, inside the $\leq 5\%$ design target
but not the $0\%$ the calibration draw shows. Both escapes are
\texttt{C\_near\_miss} ($2/30$ in that class; $0/30$ for both
\texttt{B} and \texttt{D}), and both clear $\tau_{\text{auto}}$
narrowly (promotion scores $0.79$ and $0.67$). The asymmetry is
informative rather than incidental: a verb-object permutation of a
real contract preserves almost all of the lexical and structural
evidence the deterministic profile reads, so it is the perturbation
the checks are least able to name---which is also why the
\texttt{review} tier exists. Cross-domain and swapped-contract
negatives, whose evidence profiles collapse on binding and coverage,
are separated cleanly at every threshold we tried.

\input{tables/calibration_curve}

\begin{figure}[t]
  \centering
  \includegraphics[width=0.95\columnwidth]{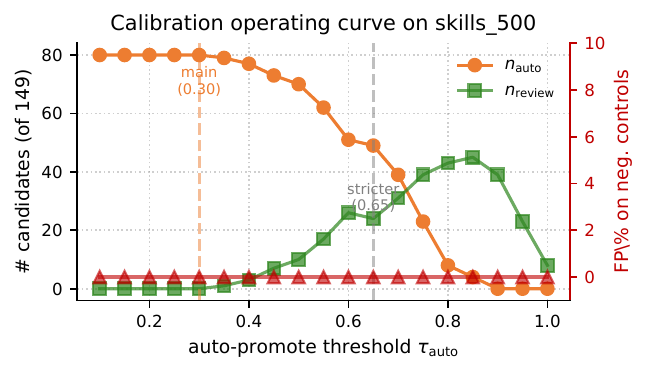}
  \caption{Calibration operating curve on \texttt{skills\_500}.
  Lowering $\tau_{\text{auto}}$ from $0.65$ to $0.30$ keeps the
  false-positive rate at $0\%$ on the calibration controls while
  admitting $31$ more real candidates ($49 \to 80$).}
  \label{fig:calibration}
\end{figure}

\autoref{tab:reject-breakdown} reports which of the four verifier
checks each non-promoted candidate failed first at the main
operating point. Binding dominates ($71\%$)---the deliberate
consequence of a high-recall proposer whose over-wide clusters the
binding check is designed to filter---followed by LLM extraction
refusals ($18\%$) and coverage mismatches ($11\%$); the risk check
contributes no first-failures on \texttt{skills\_500} (its
hard-reject fires before threshold policy but no candidate cluster
in this library introduces an undeclared unsafe sink).

\input{tables/reject_breakdown}

\section{Comparison to published GoS-paper numbers}
\label{app:vs-published}

\citet{liu2026gos} report ALFWorld reward of $89.3\%$ (Vanilla) /
$92.9\%$ (Vector) / $93.6\%$ (GoS) on $134$ games with
\texttt{gpt-5-codex}, and $27.4\% / 21.5\% / 34.4\%$ (Vanilla /
Vector / GoS) on the $87$-task SkillsBench evaluation reported by
\citet{liu2026gos} with the same model. Our absolute numbers are
substantially lower because we use the more economical
\texttt{gpt-4o-mini} for ALFWorld and a $10$-task SkillsBench subset.
We focus on \emph{relative} comparison between \textsc{gos} and
\textsc{sap} (which our setup makes apples-to-apples by holding the
agent, model, and retrieval interface fixed) rather than absolute
parity with the published numbers. The agent and retrieval interface
are identical for \textsc{gos} and \textsc{sap}; only the indexed
library content differs.

\section{Full pipeline yield}
\label{app:pipeline-yield-full}

\input{tables/pipeline_yield}

\section{Per-seed breakdown of \autoref{tab:alfworld-main}}
\label{app:multiseed-full}

\input{tables/multi_seed}

\section{Win-set diff (\textsc{sap} vs \textsc{gos}, single seed)}
\label{app:winset-diff}

\input{tables/winset_diff}

\section{Per-game outcome split (mechanism)}
\label{app:mechanism-split}

\autoref{tab:mechanism-split} gives the $3$-seed pooled per-game
\texttt{SkillRequest} and ``\texttt{Nothing happens.}'' counts split
by outcome (cited from \autoref{sec:analysis:failures}).
\autoref{tab:mechanism-split-s42} reports the single-seed
($\texttt{seed}{=}42$) version with a third ``bad acts'' column
counting actions whose verb the ALFWorld engine did not recognise as
a command (distinct from ``\texttt{Nothing happens.}'', which
indicates a recognised verb that failed in the current state).

\input{tables/mechanism_split}

\begin{table}[h]
\centering
\small
\setlength{\tabcolsep}{3pt}
\begin{tabular}{lrrrr}
\toprule
Outcome & \# & SkillReq & bad acts & ``Nothing'' \\
\midrule
\textsc{gos} -- won  & $16$  & $3.7$  & $0.19$ & $5.7$  \\
\textsc{gos} -- lost & $118$ & $13.6$ & $1.27$ & $17.7$ \\
\midrule
\method{} -- won     & $30$  & $2.7$  & $0.10$ & $3.6$  \\
\method{} -- lost    & $104$ & $9.1$  & $1.15$ & $14.7$ \\
\bottomrule
\end{tabular}
\\[2pt]{\footnotesize Per-game means; \# = games.}
\caption{Single-seed ($\texttt{seed}{=}42$) version of
\autoref{tab:mechanism-split}, adding the bad-acts column.
\method{} reduces the ``bad-acts'' rate on losing games
($1.15$ vs.\ $1.27$), the latter pinpointing the syntactic-near-miss
failure mode that the inlined action template eliminates.}
\label{tab:mechanism-split-s42}
\end{table}

\section{Per-task-type breakdown}
\label{app:per-task-type}

\input{tables/per_task_type}

\section{Retrieval-rank shift between \textsc{gos} and \method{}}
\label{app:retrieval-rank}

Rewriting skill documents changes their embeddings, so \method{}
may retrieve differently from \textsc{gos} even before content
substitution. We compare the initial task-level retrieval of the
two modes on the final post-rerank lists the pipeline records
(\textsc{gos} lists contain $4$ entries; the agent is shown the top
$3$), over $134$ paired tasks per seed. Top-$1$ agreement is
$62.7\%$ ($84/134$) on every seed; set overlap on the common list
length is $0.55 / 0.49 / 0.50$ across seeds $42/7/99$;
\textsc{gos}'s top-$1$ skill appears within \method{}'s top-$5$ in
$129 / 124 / 125$ of $134$ tasks ($93$--$96\%$); Kendall $\tau$ on
shared items is $0.50$--$0.55$ on the final lists ($0.65$ at the
embedding-seed stage before reranking). No skill is renamed or
removed by the rewrite (\texttt{skills\_500} document names are all
preserved). The shift is therefore real but local: top-$1$ changes
on $\sim 37\%$ of tasks, yet the \textsc{gos}-best skill remains
near the top of \method{}'s list in almost all tasks. The
bundle-component ablation
(\autoref{sec:results:bundle_ablation}) holds retrieval fixed
across its four content conditions and is unaffected by this shift;
the headline \method{}-vs-\textsc{gos} comparison includes both the
retrieval-shift and content effects, as any end-to-end deployment
would.

\section{Case study 1: \texttt{idx\_36}, both win, \method{} \texorpdfstring{$3.1\times$}{3.1x} faster}
\label{app:case-study-1}

Task: ``put a hot mug in a cabinet.''

\textsc{gos} wins in $22$ environment steps with $4$
\texttt{SkillRequest} cycles. The trajectory contains three
near-syntax errors: (a) the agent emits \texttt{heat mug with
microwave} before holding the mug, which the engine rejects because
the precondition (object in inventory) is not satisfied; (b) after
re-retrieving, the agent emits \texttt{use alfworld-object-locator
to find mug}, treating the retrieved skill name as an action verb;
(c) after another retrieval, the agent emits \texttt{move mug 1 to
microwave 1 (place inside)}, where the parenthetical is not parsed
by the engine's command grammar. Each error returns ``\texttt{Nothing
happens.}'' and triggers another \texttt{SkillRequest}.

\method{} wins in $7$ environment steps with a single
\texttt{READ\_SKILL} of \texttt{alfworld-heat-object-with-appliance}.
The substituted bundle places the inlined action templates
\texttt{go to \{recep\}}, \texttt{take \{obj\} from \{recep\}},
\texttt{heat \{obj\} with \{appliance\}}, \texttt{move \{obj\} to
\{recep\}} at the top of the response, so the agent emits the four
actions in order with no ``\texttt{Nothing happens.}''. The mechanism
is concrete: the bundle's structured signal supplies both the typed
contract (\emph{what} arguments to supply) and the action templates
(\emph{how} to emit them), which the \textsc{gos} prose carries in
scattered form.

\section{Case study 2: \texttt{idx\_122}, \textsc{sap}-only win}
\label{app:case-study-2}

Task: ``examine a CD with the desklamp.'' \textsc{gos} fails at
$30$ environment steps, having issued $10$ \texttt{SkillRequest}
cycles asking ``where is the CD,'' ``which drawer has the CD,''
etc.\ before finally locating the CD on \texttt{desk 2} at step
$28$---out of time to actually \texttt{use desklamp}. \method{}
wins in $11$ environment steps with \textbf{zero}
\texttt{SkillRequest} cycles: the initial retrieval guidance is
sufficient to navigate, take, move-to-desklamp, and toggle.

\section{Spurious BE call-site patterns}
\label{app:be-spurious}

The $321$ deterministic call-sites BE dropped as spurious cluster
into two dominant patterns: (i) units that share keywords with the
contract but describe a distinct procedure (a Bundled Resources
list containing the file-helper child's name, classified as an
\texttt{io-helper} call-site); (ii) units that mention the
contract's domain abstractly without invoking it (a Risk bullet
``do not delete \texttt{output.json}'' classified as a
\texttt{file-delete} call-site). Both patterns survive the
deterministic per-unit binding filter because each unit
\emph{does} contain a relevant token; only a prompt-aware LLM
seeing the full contract can tell that the unit is not an
invocation.

\section{Cross-modality dispatch}
\label{app:cross-modality}

The same pipeline runs on typed and text libraries by swapping
Stages~1--2 only: OpenAPI specs use a JSON-endpoint parser and
exact-match parameter clustering; markdown skills use YAML
frontmatter $+$ heading segmentation and embedding-based clustering.
Stages~3+ (contract extractor, four-check verifier, BE bindings, RC
cleanup, retrieval-time substitution) are unchanged. This paper's
empirical evaluation is markdown-only; cross-modality claims are
\emph{architectural}---typed-API end-to-end evaluation on Stripe /
GitHub OpenAPI is future work (\nameref{sec:limitations}).

\section{Pipeline reproducibility}
\label{sec:appendix}

\subsection{End-to-end commands}
\label{app:reproducibility}

The full pipeline runs as the following sequence of CLI invocations.
Each script takes \texttt{--lib-dir} naming the library being
processed; defaults to \texttt{results\_skills500} below.

{\scriptsize\begin{verbatim}
# 1. Parser: SKILL.md -> parents.json
python3 exp_skills500_parser.py

# 2. Candidate proposer: 5709 units -> 149 clusters
python3 exp_text_candidate_proposer.py \
    --lib-dir results_skills500

# 3. Contract extractor (LLM): 149 clusters -> drafts
python3 exp_text_contract_extractor.py \
    --lib-dir results_skills500

# 4. Verifier v2: deterministic 4-check + 3-tier policy
python3 exp_verifier_v2.py \
    --lib-dir results_skills500

# 4.5 Negative controls (for calibration)
python3 exp_negative_controls.py \
    --lib-dir results_skills500 \
    --n-per-class 10

# 4.6 Calibrate policy on negative controls
python3 exp_calibrate_policy.py \
    --lib-dir results_skills500

# 5. Skip-repair (uses calibrated policy if present)
python3 exp_skip_repair_v2.py \
    --lib-dir results_skills500

# 6. Refactor with BE bindings + RC cleanup
python3 exp_refactor_library.py \
    --lib-dir results_skills500 \
    --llm-bindings
\end{verbatim}}

\subsection{Refactored library construction}
\label{app:rebuild}

After \texttt{exp\_refactor\_library.py} produces
\texttt{refactored\_library.json} and per-parent
\texttt{*.rewritten.md} files, a separate utility constructs the
new library directory and indexes it for retrieval:
{\scriptsize\begin{verbatim}
# Build skills_500_refactored/ with all 500 original
# skills (rewritten for those parents containing
# invoke() placeholders) + 80 new child contract dirs
python3 build_refactored_skillset.py

# Index for GoS retrieval (text-embedding-3-large,
# d=3072)
cd graph-of-skills
uv run gos index \
    data/skillsets/skills_500_refactored \
    --workspace data/gos_workspace/skills_500_refactored_v1 \
    --clear
\end{verbatim}}

\subsection{ALFWorld and SkillsBench launch}
\label{app:launchers}

ALFWorld is invoked via \texttt{evaluation/alfworld\_run.py}
(supporting modes \texttt{none}, \texttt{all\_full}, \texttt{vector},
\texttt{gos}, \texttt{cos\_refactor}). The \texttt{cos\_refactor}
mode takes an extra argument:
{\scriptsize\begin{verbatim}
uv run python -m evaluation.alfworld_run \
    --mode cos_refactor \
    --refactored_library \
       results_skills500/refactored/refactored_library.json \
    --model gpt-4o-mini --use_skill \
    --gos_workspace \
       data/gos_workspace/skills_500_refactored_v1 \
    --skills_dir data/skillsets/skills_500_refactored \
    --split test --max_games 134 --max_steps 30 \
    --max_workers 4 \
    --exp_name main_134
\end{verbatim}}

SkillsBench uses Harbor + Docker + the OpenAI codex CLI; the precise
invocation for the 10-task subset is in our run scripts.

\subsection{LLM prompts}
\label{app:prompts}

\paragraph{Stage 3 contract extractor (system prompt, abridged).}
\begin{quote}
\small\ttfamily
You are extracting a typed callable contract from a cluster of
procedural units across multiple parent skills. Return STRICT JSON:
\{id, trigger, input\_schema, output\_schema, preconditions,
postconditions, side\_effects, rationale\}. If the cluster contents
are too heterogeneous to unify, return
\{\_extraction\_failed: true, reason: ...\}.
\end{quote}

\paragraph{Binding Extraction pass (system prompt, abridged).}
\begin{quote}
\small\ttfamily
You decide whether a procedural unit is genuinely a call of the
given child skill, and if so extract per-input bindings. Output:
\{should\_invoke: bool, confidence: low|medium|high, bindings:
\{name: source\_substring\}, residual\_parent\_text: ..., rationale:
...\}. For every required input you MUST provide a binding that
overlaps with the unit text, else should\_invoke=false.
\end{quote}

\paragraph{Rewrite Cleanup pass (system prompt, abridged).}
\begin{quote}
\small\ttfamily
Some passages of this rewritten parent may still describe operations
that the listed child contract already covers, or may contradict the
child's stated I/O. Rewrite such passages to use
invoke(child\_id, args). Keep parent-specific text. Do not introduce
content absent from the original. Sanity post-check: every listed
child contract must still appear at least once.
\end{quote}

\subsection{Negative-control construction}
\label{app:negcontrols}

We use three classes (the \texttt{A\_cross\_domain} class requires
multiple libraries and was not generated for the
\texttt{skills\_500} single-library run):

\begin{itemize}[leftmargin=*,topsep=2pt,itemsep=2pt]
\item \texttt{B\_same\_domain\_distinct} ($n=10$): within
\texttt{skills\_500}, pick units from different
\texttt{(verb, object)} frames and force them into a cluster.
Expectation: verifier rejects.
\item \texttt{C\_near\_miss} ($n=10$): same verb but different
object (e.g.\ \texttt{validate-email}, \texttt{validate-phone}).
Expectation: binding check fails per-parent because each parent only
has its own object word.
\item \texttt{D\_swapped\_contract} ($n=10$): a real candidate
cluster paired with a contract from a different cluster (swapped).
Expectation: coverage check fails.
\end{itemize}

All 30 negative controls were rejected at the default policy
(\autoref{tab:calibration}), confirming the policy is at least
conservative-enough to exclude these synthetic adversarial inputs.

\subsection{Per-task ALFWorld breakdown}
\label{app:alfworld-tasks}

The per-task-type win counts for the $134$-game single-seed run
($\texttt{seed}=42$) are in \autoref{tab:per-task-type}; the
game-level win-set diff between \textsc{sap} ($30$ wins) and
\textsc{gos} ($16$ wins) is in \autoref{tab:winset-diff}
($18$ \textsc{sap}-only, $4$ \textsc{gos}-only, $12$ joint wins).

\subsection{Software versions}

\texttt{gpt-4o-mini} (pipeline + ALFWorld agent),
\texttt{gpt-5-codex} (SkillsBench agent),
\texttt{text-embedding-3-large} ($d=3072$) for GoS workspace,
\texttt{text-embedding-3-small} for proposer. Harbor harness used
for SkillsBench Docker orchestration; OrbStack provides the local
Docker runtime.

%% file: tables/component_ablation.tex
\begin{table}[t]
\centering
\small
\setlength{\tabcolsep}{4pt}
\begin{tabular}{lccr}
\toprule
Configuration & BE & RC & Avg R \\
\midrule
Deterministic only & --- & --- & 0.050 (1/20) \\
+ RC cleanup & --- & \checkmark & 0.100 (2/20) \\
+ BE (Full) & \checkmark & \checkmark & \textbf{0.150 (3/20)} \\
\midrule
(ref.) \textsc{gos}, same subset & n/a & n/a & 0.150 (3/20) \\
\bottomrule
\end{tabular}
\caption{Pipeline component ablation, ALFWorld $20$-game pilot
(idx $0$--$19$). $+5$\,pp per step in the order added; BE-alone
(without RC) was not run. Full pipeline matches \textsc{gos} on the
subset and exceeds it on the full $134$
(\autoref{tab:alfworld-main}); per-pass failure modes in
\autoref{sec:results:ablation}, \autoref{sec:analysis:cleanup}.}
\label{tab:component-ablation}
\end{table}

%% file: tables/skillsbench_per_task.tex
\begin{table*}[t]
\centering
\small
\setlength{\tabcolsep}{4pt}
\begin{tabular}{lcccrrrrrr}
\toprule
Task & \textsc{gos} R & \method{} R & & \textsc{gos} in & \method{} in & $\Delta$ in & \textsc{gos} out & \method{} out & $\Delta$ out \\
\midrule
\texttt{citation-check}         & $0$ & \textbf{$1$\,\checkmark} & & $0.92$M & $0.66$M & $\mathbf{-28\%}$ & $14.2$k & $8.0$k  & $\mathbf{-44\%}$ \\
\texttt{court-form-filling}     & $0$ & $0$\,$\triangle$         & & $2.32$M & $0.95$M & $-59\%$ & $25.7$k & $11.9$k & $-54\%$ \\
\texttt{data-to-d3}             & $0$ & $0$\,$\triangle$         & & $1.00$M & $0.84$M & $-15\%$ & $23.2$k & $14.6$k & $-37\%$ \\
\texttt{dialogue-parser}        & $1$ & $1$                       & & $0.58$M & $1.31$M & $+125\%$ & $28.0$k & $24.6$k & $-12\%$ \\
\texttt{edit-pdf}               & $0$ & $0$                       & & $0.92$M & $0.32$M & $\mathbf{-65\%}$ & $18.8$k & $8.5$k  & $-55\%$ \\
\texttt{exceltable-in-ppt}      & $0$ & $0$                       & & $0.38$M & $0.33$M & $-14\%$ & $7.5$k  & $5.0$k  & $-34\%$ \\
\texttt{fix-build-agentops}     & $0$ & $0$                       & & $18.24$M& $12.19$M& $-33\%$ & $55.2$k & $32.4$k & $-41\%$ \\
\texttt{fix-build-google-auto}  & $0$ & $0$                       & & $8.90$M & $2.00$M & $\mathbf{-78\%}$ & $30.7$k & $9.3$k  & $-70\%$ \\
\texttt{offer-letter-generator} & $0$ & $0$                       & & $0.36$M & $0.07$M & $\mathbf{-82\%}$ & $5.4$k  & $1.4$k  & $-74\%$ \\
\texttt{powerlifting-coef-calc} & $1$ & $1$                       & & $0.34$M & $0.55$M & $+59\%$ & $10.0$k & $18.0$k & $+80\%$ \\
\midrule
\textbf{TOTAL}                  & $\mathbf{2}$ & $\mathbf{3}$ & & $\mathbf{33.96}$M & $\mathbf{19.21}$M & $\mathbf{-43\%}$ & $\mathbf{218.8}$k & $\mathbf{133.6}$k & $\mathbf{-39\%}$ \\
\bottomrule
\end{tabular}
\caption{Per-task breakdown of \autoref{tab:skillsbench-main} on
the $10$-task SkillsBench sample. ``R'' = binary task reward
($1$ = pass verifier, $0$ = fail). $\checkmark$ = \method{}-only win
(\textsc{gos} fails); $\triangle$ = \method{} run failed mid-execution
(\texttt{NonZeroAgentExitCodeError}, counted as $0$ reward);
input tokens were already $-59\%$ / $-15\%$ below \textsc{gos}
at the crash, so retrieval was not the bottleneck, but token
count alone does not establish bundle correctness.
\method{} reduces input tokens on $8/10$ tasks (range $-14\%$
to $-82\%$). Two tasks see \method{} use more tokens
(\texttt{dialogue-parser}, \texttt{powerlifting-coef-calc}): both
are tasks where \emph{both} modes win, and \method{} writes more
code to reach the same correct answer. Aggregated: $-43\%$ input,
$-39\%$ output tokens (matches \autoref{tab:skillsbench-main}).}
\label{tab:skillsbench-per-task}
\end{table*}

%% file: tables/calibration_curve.tex
\begin{table}[t]
\centering
\small
\setlength{\tabcolsep}{4pt}
\begin{tabular}{rrrrrl}
\toprule
$\tau_{\text{auto}}$ & $\tau_{\text{rev}}$ & $n_{\text{auto}}$ &
$n_{\text{rev}}$ & FP\% & note \\
\midrule
0.30 & 0.10 & \textbf{80} & 0  & \textbf{0.0} & main \\
0.35 & 0.15 & 79 & 1  & 0.0 & \\
0.40 & 0.20 & 77 & 3  & 0.0 & \\
0.45 & 0.25 & 73 & 7  & 0.0 & \\
0.50 & 0.30 & 70 & 10 & 0.0 & \\
0.55 & 0.35 & 62 & 17 & 0.0 & \\
0.60 & 0.40 & 51 & 26 & 0.0 & \\
0.65 & 0.35 & 49 & 24 & 0.0 & stricter (App.~H) \\
0.70 & 0.50 & 39 & 31 & 0.0 & \\
0.75 & 0.55 & 23 & 39 & 0.0 & \\
0.80 & 0.60 & 8  & 43 & 0.0 & \\
0.85 & 0.65 & 4  & 45 & 0.0 & \\
0.90 & 0.70 & 0  & 39 & 0.0 & \\
\bottomrule
\end{tabular}
\caption{Calibration operating curve on \texttt{skills\_500}.
Real candidates $n=149$; calibration negatives $n=30$ (B+C+D, $10$
each). FP\% is the false-positive rate on those controls; a held-out
draw of $90$ puts the selected point at $2.2\%$
(\autoref{app:calibration}).
Promotion count is monotone non-decreasing as thresholds drop because
no row violates FP $\leq 5\%$.}
\label{tab:calibration}
\end{table}

%% file: tables/reject_breakdown.tex
\begin{table}[t]
\centering
\small
\setlength{\tabcolsep}{4pt}
\begin{tabular}{lr}
\toprule
First-failed check (non-promoted, $n=100$) & Count \\
\midrule
\texttt{binding}              & 71 \\
\texttt{extraction\_failed}   & 18 \\
\texttt{coverage\_recall}     & 11 \\
\midrule
Terminal tier (all $n=149$) & \\
\midrule
\texttt{auto\_promote}        & 49 \\
\texttt{review}               & 30 \\
\texttt{reject}               & 70 \\
\bottomrule
\end{tabular}
\caption{Verifier-layer breakdown on \texttt{skills\_500} at the
main operating point. \texttt{extraction\_failed}: the LLM refused
to produce a contract; \texttt{binding}: per-parent input evidence
missing; \texttt{coverage\_recall}: contract tokens absent from
parent text. Binding dominates first failures ($71\%$) by design:
the candidate proposer is deliberately high-recall, so over-wide
clusters are the most common failure and the binding check is their
designated filter. Each layer catches a distinct class; removing
any one layer would convert its rejections into silent false
positives.}
\label{tab:reject-breakdown}
\end{table}

%% file: tables/pipeline_yield.tex
\begin{table}[t]
\centering
\small
\setlength{\tabcolsep}{3pt}
\begin{tabular}{lrr}
\toprule
Stage & Input & Output \\
\midrule
Parser & 500 docs & 5{,}709 units \\
Candidate proposer & 5{,}709 & 149 clusters \\
Contract extractor (LLM) & 149 & 131 + 18 refusals \\
Verifier (main policy) & 149 & 49 / 30 / 70$^{*}$ \\
Call-site detect (det.) & 49 & 791 sites \\
BE bindings (LLM) & 791 & 470 (59\%) \\
Refactor + RC & 292 parents & 291 (99.7\%) \\
\bottomrule
\end{tabular}
\caption{Pipeline yield on \texttt{skills\_500} at the main
operating point $(\tau_{\text{auto}},\tau_{\text{rev}})=(0.65,0.35)$
used for all main experiments.
$^{*}$\texttt{auto\_promote} / \texttt{review} / \texttt{reject}.
A looser calibrated point $(0.30,0.10)$ admits $80/149$ at the same
$0\%$ FP rate on the calibration negatives; that variant is reported in
\autoref{app:default-policy}. Pipeline LLM cost:
$\sim$\$$1.3$ at \texttt{gpt-4o-mini} list prices.}
\label{tab:pipeline-yield}
\end{table}

%% file: tables/multi_seed.tex
\begin{table*}[t]
\centering
\small
\setlength{\tabcolsep}{4pt}
\begin{tabular}{llrrrrr}
\toprule
Mode & Metric & seed$=42$ & seed$=7$ & seed$=99$ & mean $\pm$ std & pooled ($402$) \\
\midrule
\multicolumn{7}{l}{\emph{Reward.}} \\
\textsc{gos}                  & wins / $134$  & $16$    & $16$    & $15$    & $15.67 \pm 0.58$ & $47$ \; ($11.7\%$) \\
\textbf{\textsc{sap} (\method{})} & wins / $134$ & $\mathbf{30}$ & $\mathbf{30}$ & $\mathbf{22}$ & $\mathbf{27.33 \pm 4.62}$ & $\mathbf{82}$ \; ($\mathbf{20.4\%}$) \\
\addlinespace[3pt]
\multicolumn{7}{l}{\emph{Input tokens per game (k).}} \\
\textsc{gos}                  & input / g  & $247.1$ & $248.8$ & $247.5$ & $247.8 \pm 0.9$ & $247.8$ \\
\textbf{\textsc{sap}}         & input / g  & $\mathbf{177.8}$ & $\mathbf{187.3}$ & $\mathbf{209.0}$ & $\mathbf{191.4 \pm 15.9}$ & $\mathbf{191.4}$ \\
                              & $\Delta$\,(\%) & $-28.0$ & $-24.7$ & $-15.6$ & $\mathbf{-22.8 \pm 6.4}$ & $-22.8$ \\
\addlinespace[3pt]
\multicolumn{7}{l}{\emph{Output tokens per game.}} \\
\textsc{gos}                  & output / g & $1938$ & $1962$ & $1905$ & $1935 \pm 29$ & $1935$ \\
\textbf{\textsc{sap}}         & output / g & $\mathbf{1525}$ & $\mathbf{1587}$ & $\mathbf{1687}$ & $\mathbf{1599 \pm 82}$ & $\mathbf{1599}$ \\
                              & $\Delta$\,(\%) & $-21.3$ & $-19.1$ & $-11.5$ & $\mathbf{-17.3 \pm 5.1}$ & $-17.3$ \\
\addlinespace[3pt]
\multicolumn{7}{l}{\emph{LLM calls per game.}} \\
\textsc{gos}                  & calls / g  & $40.2$ & $39.1$ & $39.8$ & $39.7 \pm 0.5$ & $39.7$ \\
\textbf{\textsc{sap}}         & calls / g  & $\mathbf{32.5}$ & $\mathbf{33.9}$ & $\mathbf{35.4}$ & $\mathbf{33.9 \pm 1.5}$ & $\mathbf{33.9}$ \\
                              & $\Delta$\,(\%) & $-19.1$ & $-13.4$ & $-11.1$ & $\mathbf{-14.5 \pm 4.1}$ & $-14.5$ \\
\bottomrule
\end{tabular}
\caption{Per-seed breakdown of \autoref{tab:alfworld-main}.
\method{} wins more games than \textsc{gos} on every seed in both
reward and tokens. \textsc{gos}'s per-game input is near-constant
across seeds ($\pm 0.4\%$); \method{}'s drops more on seeds where it
wins more (won games close out before the
$\texttt{max\_steps}{=}30$ budget), shrinking from $-28\%$ on
$\texttt{seed}{=}42$ to $-16\%$ on $\texttt{seed}{=}99$. The reward
gain and the token saving co-occur on every seed.}
\label{tab:multiseed-full}
\end{table*}

%% file: tables/winset_diff.tex
\begin{table}[t]
\centering
\small
\begin{tabular}{lr}
\toprule
Win group & \# games \\
\midrule
\textsc{sap}-only (\textsc{gos} lost)  & \textbf{18} \\
\textsc{gos}-only (\textsc{sap} lost)  & 4 \\
both win                               & 12 \\
both lose                              & 100 \\
\bottomrule
\end{tabular}
\caption{Win-set diff between \method{} and \textsc{gos} on the
ALFWorld $134$-game unseen split (single seed $=42$). \method{}
wins $18$ games \textsc{gos} loses while losing only $4$ that
\textsc{gos} wins; both win on a shared $12$. Paired McNemar exact
on the $22$ discordant pairs gives $p=0.0043$. The $100$ both-lose
games are tasks that exceed the action budget at
$\texttt{max\_steps}=30$ for either configuration with
\texttt{gpt-4o-mini}; \autoref{sec:limitations} discusses model
ceiling.}
\label{tab:winset-diff}
\end{table}

%% file: tables/mechanism_split.tex
\begin{table}[t]
\centering
\small
\setlength{\tabcolsep}{4pt}
\begin{tabular}{llrrr}
\toprule
Mode & Outcome & games & SkillReq / g & ``Nothing'' / g \\
\midrule
\textsc{gos}          & won  & $47$  & $3.6$  & $4.3$  \\
\textsc{gos}          & lost & $355$ & $13.2$ & $17.5$ \\
\addlinespace[1pt]
\textbf{\textsc{sap}} & won  & $\mathbf{82}$  & $3.2$  & $3.7$  \\
\textbf{\textsc{sap}} & lost & $320$ & $9.2$  & $15.3$ \\
\bottomrule
\end{tabular}
\caption{Per-game means split by outcome, pooled across $3$ seeds.
Within either mode, losing games issue $\sim 4\times$ more
\texttt{SkillRequest}s and $\sim 4\times$ more ``\texttt{Nothing
happens.}'' observations than winning games. The loop is the unit of
failure (the asymmetry holds inside each mode), not a mode-specific
artifact. \method{} wins more games \emph{and} reduces both counts on
both populations: \method{}-won games use $\sim 10\%$ fewer
\texttt{SkillRequest}s than \textsc{gos}-won games, and \method{}-lost
games use $\sim 30\%$ fewer than \textsc{gos}-lost games---first
retrieval more often suffices.}
\label{tab:mechanism-split}
\end{table}

%% file: tables/per_task_type.tex
\begin{table*}[t]
\centering
\small
\begin{tabular}{lrrrr}
\toprule
Task type & games & \textsc{gos} & \textbf{\textsc{sap} (\method{})} & diff \\
\midrule
\texttt{pick\_clean\_then\_place}   & 31 & 5 (16.1\%) & 7 (22.6\%)  & $+2$ \\
\texttt{pick\_and\_place\_simple}   & 24 & 3 (12.5\%) & \textbf{7 (29.2\%)}  & $+4$ \\
\texttt{pick\_heat\_then\_place}    & 23 & 2 (8.7\%)  & 4 (17.4\%)  & $+2$ \\
\texttt{pick\_cool\_then\_place}    & 21 & 2 (9.5\%)  & \textbf{5 (23.8\%)} & $+3$ \\
\texttt{look\_at\_obj\_in\_light}   & 18 & 3 (16.7\%) & 4 (22.2\%)  & $+1$ \\
\texttt{pick\_two\_obj\_and\_place} & 17 & 1 (5.9\%)  & 3 (17.6\%)  & $+2$ \\
\midrule
all (single seed $=42$)             & 134 & 16 (11.9\%) & \textbf{30 (22.4\%)} & $+14$ \\
\bottomrule
\end{tabular}
\caption{Per-task-type breakdown on ALFWorld $134$-game (single seed
$=42$, \texttt{gpt-4o-mini}). \method{} wins on \emph{every} task
type, with $+2$ to $+4$ absolute wins per type. The largest absolute
gains are on \texttt{pick\_and\_place\_simple} ($+4$) and
\texttt{pick\_cool\_then\_place} ($+3$), both of which require a
multi-step navigate-take-(modify-)place sequence that benefits from
the inlined concrete action templates described in
\autoref{sec:method:retrieval}.}
\label{tab:per-task-type}
\end{table*}